\documentclass[11pt,a4paper,twoside,pdf]{article}

\usepackage{array}
\usepackage{amsmath}
\usepackage{upgreek}
\usepackage{marvosym}
\usepackage{epsfig}
\usepackage{graphics}
\usepackage{float}
\usepackage{amsfonts}
\usepackage{xspace}
\usepackage{booktabs}
\usepackage{tabularx}
\usepackage{xtab}
\usepackage{feynmp-auto}
\numberwithin{equation}{section}
\usepackage{braket}
\usepackage{titlesec}
\usepackage[skins,breakable]{tcolorbox}
\usepackage{feynmp}
\usepackage[english]{babel} 
\addto\captionsenglish{}

\usepackage{soul}
\titleformat{\section}{\sffamily \Large \bfseries \color{black!20!lightviolet}}{\thesection.\enspace}{0.5em}{}[{\titlerule[2pt]}\vspace*{4pt}]
\titleformat{\subsection}{\sffamily \large \bfseries \color{white!5!lightviolet}}{\thesubsection.}{0.5em}{}
\titleformat{\subsubsection}{\sffamily \normalsize \bfseries \color{white!10!lightviolet}}{\thesubsubsection.}{0.5em}{}

\makeatletter
\@ifpackageloaded{xcolor}{}{
	\usepackage{xcolor}
}
\makeatother

\definecolor{lightviolet}{rgb}{0.454902,0,0.313725}
\definecolor{colorblue}{rgb}{0.0898438,0.375,0.507813}
\definecolor{darkviolet}{rgb}{0.314,0.086,0.29}
\definecolor{greentwo}{RGB}{85,132,142}

\usepackage[backend=biber,style=numeric-comp,sorting=none]{biblatex}
\renewbibmacro{in:}{}
\DeclareFieldFormat{pages}{#1}
\renewbibmacro{in:}{}

\usepackage{fancyhdr}
\usepackage[a4paper,bottom=3cm,top=2.5cm,left=2.0cm,right=2.0cm,head=0mm,dvipdfm]{geometry}
\usepackage{verbatim}
\usepackage{latexsym}
\usepackage{amssymb}
\usepackage{graphicx}
\usepackage{arydshln}
\usepackage{wasysym}
\usepackage{adjustbox}
\usepackage{easybmat}
\usepackage{bbm}
\usepackage{bm}
\usepackage{multirow} 
\usepackage{rotating}
\usepackage{mathtools}
\usepackage[inline]{enumitem}
\usepackage{lscape}
\usepackage[normalem]{ulem}
\usepackage{colortbl}
\usepackage{pstricks}
\usepackage[font=small,labelfont=bf]{caption}
\usepackage{subcaption}
\usepackage{lineno}
\usepackage{csquotes}
\usepackage{indentfirst}
\usepackage{hyphenat}
\usepackage{blindtext}
\usepackage{authblk}
\usepackage{mathptmx}
\usepackage{slashed}
\usepackage{url}
\usepackage{tocloft}
\usepackage[T1]{fontenc}
\usepackage{listings}
\usepackage[sc]{mathpazo}
\usepackage{anyfontsize}

\usepackage{hyperref}
\hypersetup{
    colorlinks=true,
    linkcolor=black,
    urlcolor=colorblue,
    citecolor=lightviolet
}

\newcolumntype{C}{>{\centering\arraybackslash} m{2.5cm} }
\newcolumntype{Y}{>{\centering\arraybackslash}X}

\newcommand{\tool}{\texttt{mosca}}
\newcommand{\mathematica}{\textsc{\texttt{Mathematica}}}

\usepackage{mmacells}

\mmaSet{
  morefv={gobble=2},
  linklocaluri=mma/symbol/definition:#1,
  morecellgraphics={yoffset=1.9ex},
  moredefined={GenerateGeneralKinematics, PropagatorAttributes, AllPropagatorAttributes, AmplitudeComputation, AmplitudeMatching, MPhysSymbol,OutputFormat,Replacements, RecoverFullAmplitude, MassReduction, FormatWC, ModelName, ProcessList, FindIsomorphisms, Patch2ptFunction, Patch2mosca, Bare2PhysMass, PatchMasses,EFTSeries, VisibleOrder, VisibleModel, PatchModelsOnly, FAPatch, CouplingRename, WriteFeynArtsOutput, FR, Loop, SeparateCrossings, ReplaceBareMass, ReplacePhysMass, CouplingOrderSust, WClist, KinematicConfigurations, Sorted, Values, MDot, pslash, ReplaceKinematics, CreateTopologies,  InsertFields, InsertionLevel, Model, GenericModel, FCFAConvert, CreateFeynAmp, IncomingMomenta, DropSumOver, FeynAmpDenominatorExplicit, SUNSimplify, SUNFJacobi, PolarizationInfo, Momenta, DefineWC,EFTOrder,ComplexParameter,DefineAllWCs,FileNameJoin,TrueEFTOrder,WC, inv, RenameWC, Output, F, V, S, ExcludeParticles, RedBasis, StopAt, TopologyList, mphys2, Z, Prop, Paint, ColumnsXRows, ExplicitEFTOrder, SolvePerturbative, FeynAmpDenonimator, Bare2PhysMass, GammaDenominator, DiracGamma, RecalcPropAttributes, DiagramComputation, Reduction, JoinOrderWC
  }
}

\mmaDefineMathReplacement{`}{\`}
\mmaDefineMathReplacement{\$}{\$}

\newcommand{\partL}{\textlbrackdbl}
\newcommand{\partR}{\textrbrackdbl}

\newcommand{\tab}{\hspace{12pt}}

\newcommand{\txtbox}[3]{

{\small
\begin{tcolorbox}[breakable, colback=white,middle=0mm,boxsep=2mm,colframe=white!30!black,colbacktitle=white!50!black,fonttitle=\bfseries,title=\normalsize{#2}]
\setlength\linenumbersep{-.1cm} 
\setcounter{linenumber}{#1}
\begin{internallinenumbers}
\ttfamily
#3
\end{internallinenumbers}
\end{tcolorbox}
}

}

\newenvironment{functiondef}
{
\begin{tcolorbox}[enhanced, colback=white, frame style={left color=colorblue, right color=greentwo}, sharp corners=downhill, arc=3mm, boxrule=.8mm]
\centering
}
{
\end{tcolorbox}
}

\newenvironment{mathematicaNotebook}
{
\begin{tcolorbox}[breakable, colback=black!2!white, sharp corners, left=0mm, colframe=white!40!black, boxrule=0.3mm]
\small
}
{
\end{tcolorbox}
}

\definecolor{whitish}{RGB}{240, 231, 213}

\newcommand{\vcenteredinclude}[2]{\begingroup
	\setbox0=\hbox{\includegraphics[width=#1]{#2}}%
	\parbox{\wd0}{\box0}\endgroup}

\begin{document}

\begin{titlepage}
    \centering
    \vspace*{1.5cm}  
    {\huge Automation of a \textcolor{lightviolet}{\textbf{M}}atching \textcolor{lightviolet}{\textbf{O}}n-\textcolor{lightviolet}{\textbf{S}}hell \textcolor{lightviolet}{\textbf{Ca}}lculator \par}
       \vspace{1cm} 
    \begin{figure}[ht]
        \centering
        \includegraphics[width=0.5\linewidth]{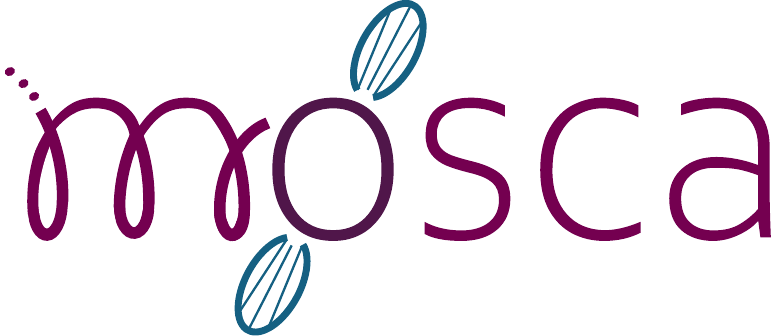}
        \label{fig:enter-label}
    \end{figure}
    \vspace{1cm}  

    {\Large Javier López Miras and Fuensanta Vilches \par}
    \vspace{0.5cm}  

    {\textit{Departamento de Física Teórica y del Cosmos} \\
    \textit{Campus de Fuentenueva, Universidad de Granada, E-18071 Granada, Spain} \par}
    \vspace{0.5cm}

    {May, 2025 \par}
    \vspace{1cm}

\begin{minipage}{0.9\textwidth}  
    We introduce \tool, a \mathematica\ package designed to facilitate on-shell calculations in effective field theories (EFTs). This initial release focuses on the reduction of Green's bases to physical bases, as well as transformations between arbitrary operator bases. The core of the package is based on a diagrammatic on-shell matching procedure, grounded in the equivalence of physical observables derived from both redundant and non-redundant Lagrangians. \tool\ offers a complete set of tools for performing basis transformations, diagram isomorphism detection, numerical substitution of kinematic configurations, and symbolic manipulation of algebraic expressions. Planned future developments include extension to one-loop computations, thus providing support for EFT renormalization directly in a physical basis and automated computation of one-loop finite matching, including contributions from evanescent operators. The package, along with example notebooks and documentation, is available at:  \url{https://gitlab.com/matchingonshell/mosca}.

\end{minipage}
\end{titlepage}

{
  \hypersetup{linkcolor=black}
  \tableofcontents
}
\newpage
\section{Introduction}
\label{sect: intro}

Effective field theories (EFTs), which have been used since the very foundations of quantum field theory, have become one of the clearest directions to choose in beyond the Standard Model (SM) research. With the rise in prominence of this field, more and more ultraviolet (UV) models are to be matched to their respective EFTs, \textit{e.g.} in order to study their physical implications at low energy levels (see~\cite{Hartland:2019bjb,Aoude:2020dwv,Dawson:2020oco,Anisha:2020ggj,Falkowski:2020pma,Ellis:2020unq,Ethier:2021bye,deBlas:2022ofj} for some applications).  Furthermore, after an initial understanding of their effects based on rough approximations, studies have shifted towards the inspection of the consequences when the full set of effective operators is considered. Well-known complete bases for some common EFTs in the literature can be found in~\cite{Grzadkowski:2010es,delAguila:2008ir,Aparici:2009fh,Bhattacharya:2015vja,Liao:2016qyd,Aebischer:2024csk,Jenkins:2017jig,Bischer:2019ttk,Chala:2020vqp,Li:2020lba,Li:2020wxi,Gripaios:2016xuo,Chala:2020wvs,Bauer:2020jbp,Grojean:2023tsd,Murphy:2020rsh,Murphy:2020cly,Li:2020tsi,Li_2021,Li:2021tsq,Li:2022tec,Harlander:2023ozs}.  This includes accounting for their precise coupling dependencies and exploring the impact of extending the EFT to higher dimensions. Hence, a consistent, reliable and efficient matching procedure is essential in order to progress with the EFT program.

In the functional approach to matching, a functional integration over the heavy states allows the extraction of the local contributions that are relevant for the description of the low-energy dynamics. These local contributions correspond to operators which are typically not in a physical basis and require reduction via field redefinitions~\cite{Arzt:1993gz,Criado:2018sdb}, which can be tedious at higher orders but can be automated with tools like \texttt{Matchete}~\cite{Fuentes-Martin:2022jrf}. On the other hand, the diagrammatic approach (automated in \texttt{MatchMakerEFT}~\cite{Carmona:2021xtq}) offers two distinct implementation strategies. The first involves performing an off-shell matching, where off-shell Green's functions are computed in both the UV model and the EFT; and their difference is matched by adjusting the Wilson coefficients (WCs) of the local operators in the EFT. Although this first method is advantageous due to the smaller number of diagrams (only the 1-light-particle-irreducible should be considered), it still requires a Green's basis with redundant operators that must be reduced. The second strategy, in contrast, avoids the need for a Green's basis and allows for the direct use of physical operators by performing an on-shell matching between both theories~\cite{Georgi:1991ch,Li:2023edf,DeAngelis:2023bmd}. However, it faces challenges due to the larger number of diagrams and the non-local nature of light bridges, making it difficult to manage analytic cancellations. In~\cite{chala2024efficientonshellmatching} it was showed the implementation of a numerical (but exact) approach that efficiently addresses these challenges, proving the potential of on-shell matching.

In this work, we adopt this latter approach and automate it in a \mathematica\ code. For the sake of completeness, we shall first summarize here the algorithm for on-shell matching, as presented in~\cite{chala2024efficientonshellmatching}:
\begin{enumerate}
    \item Determination of the physical masses and residues that fix the on-shell conditions: The light propagators in the UV theory read

         \begin{equation}
        \vcenter{\hbox{\includegraphics[width=3cm]{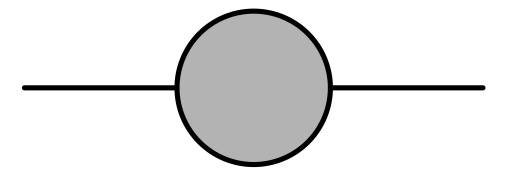}}} = \frac{i}{p^2-m^2-\Pi(p^2)} \approx \frac{iZ}{p^2-m^2_{phys}} ~,
         \label{eq: modified prop}
        \end{equation}

        \noindent
        where $\Pi(p^2)$ is the 1-particle-irreducible (1PI) contribution to the 2-point functions. In the last equality, we approximate the propagator in the vicinity of the physical mass, which corresponds to the pole of the 2-point function. The parameter $Z$ represents the residue at the physical pole and can be determined by expanding $\Pi(p^2)$ near $p^2=m^2_{phys}$ in Eq. (\ref{eq: modified prop}). Through this expansion, we find that $Z=\left(1-\Pi'(m^2_{phys})\right)^{-1}$. In the EFT, there are no 1PI corrections to the 2-point function so $m^2_{phys}=m^2$ and $Z=1$.

    \item Calculation of amplitudes: Evaluate all necessary connected, amputated amplitudes, up to the correct order in mass dimension, in both the UV model and the EFT. Then, substitute into the amplitudes the physical mass and complete propagators, and multiply by a factor $\sqrt{Z}$ for each external leg, according to the LSZ reduction formula.

    \item Solve for the WCs: Start with the simplest amplitudes (fewest external particles) and solve for the EFT WCs using on-shell randomly generated kinematic configurations. At tree level, matching equates the tree-level amplitudes of the EFT and the UV theory. 
\end{enumerate}

Furthermore, the generation of on-shell numerical kinematics with rational values is based on the algorithm presented in~\cite{DeAngelis:2022qco}. This method offers two key advantages: first, the use of rational kinematics ensures exact solutions; second, it allows for the enforcement of independent symbolic masses for each particle, a crucial feature to capture the mass dependence in the matching procedure. 

The implementation of such algorithm, which makes use of the so-called \textit{spinor-helicity} formalism \cite{AccettulliHuber:2023ldr,Arkani-Hamed:2017jhn,Cheung:2009dc,Maitre:2007jq}, can be found in a \mathematica\ package in \url{https://github.com/StefanoDeAngelis/SpinorHelicity/tree/main/Wolfram/NumericalKinematics.wl}. This package outputs both the set of physical momenta and the corresponding associated spinors. However, for processes involving massless particles of spin-1, we also require polarization vectors. Using the spinor-helicity formalism, we derive polarization vectors as follows:
\begin{equation*}
    \varepsilon_+^\nu=\frac{1}{\sqrt{2}}\frac{\lambda^\alpha \widetilde\mu^{\dot{\alpha}}\sigma^\nu_{\alpha \dot{\alpha}}}{ \lambda_\beta\mu^\beta} \quad \text{and} \quad \varepsilon_-^\nu=\frac{1}{\sqrt{2}}\frac{\mu^\alpha \widetilde\lambda^{\dot{\alpha}}\sigma^\nu_{\alpha \dot{\alpha}}}{\widetilde\lambda_{\dot{\beta}} \widetilde\mu^{\dot{\beta}}}~,
\end{equation*}
where ($\mu,\widetilde\mu$) are auxiliary spinors, ($\lambda,\widetilde\lambda$) are the spinors corresponding to the momentum of the spin-1 particle, and $\sigma^\nu =(1_{2\times 2},\,\sigma^I)$, with $\sigma^I$ the Pauli matrices. These expressions inherently satisfy the transversality conditions, ensuring they represent physical polarization states. The spinors lower and rise their indices as $\lambda^\alpha \equiv  \epsilon^{\alpha \beta}\lambda_\beta$, with $\epsilon^{\alpha \beta}$ the totally antisymmetric tensor with $\epsilon^{12}=1$. Furthermore, for spin-$\frac{1}{2}$ particles, spinors satisfying the Dirac equation are also required. In the case of incoming massless fermions, the relationship between left- and right-handed Dirac spinors and the Weyl spinor basis is established through the following relations:
\begin{equation*}
    P_L u(p) = \begin{pmatrix}
                \widetilde\lambda_{\dot{\alpha}}  \\
                0 
                \end{pmatrix} ~, \quad 
    P_R u(p) = \begin{pmatrix}
                0  \\
                \lambda^\alpha
                \end{pmatrix}  ~, \quad 
    \overline{v}(p) P_R= \begin{pmatrix}
                \widetilde\lambda^{\dot{\alpha}}  \\
                0
                \end{pmatrix} \quad \text{and} \quad 
    \overline{v}(p) P_L=\begin{pmatrix}
                0  \\
                \lambda_\alpha
                \end{pmatrix} ~,
\end{equation*}
where ($\lambda,\widetilde\lambda$) represent the spinors associated with the four-momentum $p$ of the spin-1/2 particle. Once all the kinematic structures are generated, we substitute them into the amplitudes, incorporating the numerical values of the gamma matrices in the Dirac representation. This yields a final expression where the WCs---the couplings of local operators in the effective Lagrangian of the EFT---are the only remaining unknowns.

\begin{figure}[ht]
    \centering
    \includegraphics[width=0.95\linewidth]{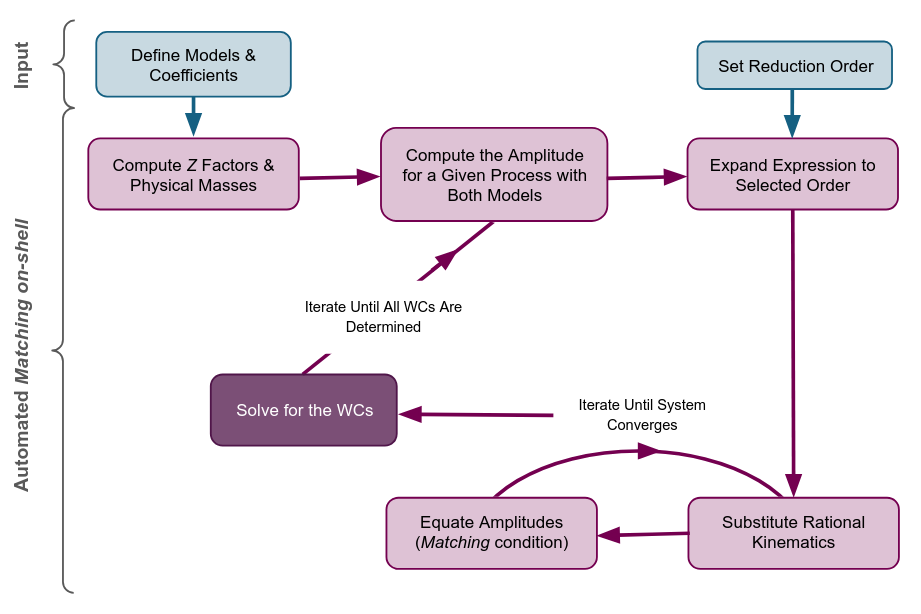}
    \caption{Schematic representation of \tool \space roadmap.}
    \label{fig: workflow}
\end{figure}

After having set the details of the on-shell matching procedure, we turn now to the scope of this manual: \tool. The \mathematica\ code \tool\ (\texttt{Matching On-Shell Calculator}) is engineered to proceed with the aforementioned routine to implement on-shell matching. The general lines of the workflow of \tool \space are summarized in Figure \ref{fig: workflow}.

Until now, the discussion of the on-shell matching procedure has focused on its traditional application in the construction of an EFT that reproduces the low-energy behavior of a more fundamental theory. However, in this initial version, \tool\ sets the usual matching of UV models aside and focuses on the matching of different Lagrangian representations of the same theory. The foundational principle of on-shell matching relies in explicitly ensuring that two Lagrangians provide identical S-matrix elements for all computable processes, up to the corresponding order in the EFT expansion. In the context of basis reduction, this same logic can be applied, since this process involves merely redefining the Lagrangian representation of the theory without altering the underlying physics. In this reinterpretation, instead of comparing the Lagrangians of a UV theory and an EFT, we compare two Lagrangian representations of the same theory expressed in terms of different operator bases.   As argued more formally in~\cite{chala2024efficientonshellmatching}, a tree-level on-shell matching can be used to relate the WCs of a Lagrangian $\mathcal{L}_1$ with respect to other $\mathcal{L}_2$, as long as there exists a local redefinition of the fields transforming $\mathcal{L}_1$ into $\mathcal{L}_2$.

As previously noted, computations with EFTs (\textit{e.g.} renormalization group evolutions or finite matching) often require a (on-shell redundant) Green's basis which is later to be reduced to the much smaller and manageable physical basis. Traditionally this has been done by simply replacing the equations of motion (EOMs) into the redundant Lagrangian \cite{Kluberg-Stern:1975ebk,Grosse-Knetter:1993tae,Wudka:1994ny,Manohar:2018aog}. However, this is known to be just a mere approximation beyond the linear order in the EFT cut-off $\Lambda$ \cite{Criado:2018sdb}. The right way to do this is via field redefinitions \footnote{Other methods such as modified equations of motions can also be considered \cite{Banerjee:2022thk}.} which, while straightforward in principle, can turn very tedious and cumbersome. Thus, a simple tree-level on-shell matching between the Lagrangians expressed in terms of some bases is a promising alternative to automate such process. In light of this, \tool\ is designed to facilitate the process of reducing Lagrangians by performing a matching based on on-shell techniques.

This paper serves as a concise introduction to \tool, providing an overview of the package's structure, main functions and practical examples. With this objective, the rest of this article is organized as follows: First, Section~\ref{sect: starting} presents the installation process and an overview of the structure of the package, including key aspects and guidance for setting up the input. Next, Section~\ref{sect: guide} introduces the main objects and functions, focusing on each part of the on-shell matching procedure. Finally, Section~\ref{sect:using mosca} is dedicated to discuss several aspects related to the use of \tool. In Section~\ref{sect: conclusions} we conclude.

\section{Getting Started: Installation and Package Overview}
\label{sect: starting}
\subsection{First time running \tool}

The \tool \space package can be downloaded from the GitLab repository:

\begin{center} \url{https://gitlab.com/matchingonshell/mosca} \end{center}

\noindent
The \tool \space folder includes the main program (\texttt{mosca.m}) along with all the essential tools needed to run the package. Among the different files, you will find the \texttt{ExternalPackages/} directory, which includes the following dependencies:

\begin{itemize}
    \item \texttt{FeynCalc}~\cite{Brambilla:2020fla}: A \mathematica \space library for symbolic manipulation of Feynman integrals and expressions in quantum field theory. 
    \item \texttt{FeynArts}~\cite{Hahn:2000kx}: A package for generating Feynman diagrams, which is included within the \texttt{FeynCalc} folder.
    \item \texttt{NumericalKinematics}: The package that provides the framework necessary for defining and managing kinematic variables.
\end{itemize}

\noindent
The \texttt{NumericalKinematics} package is used by \texttt{KinematicSubstitution}, another package located in a folder of the same name. This package constructs all the necessary physical kinematic configurations, as described in Section~\ref{sect: intro}.

The minimum requirement for running \tool\ is having installed \mathematica\ (version 8 or later, although it has been only tested from version 12 onward). The necessary \texttt{FeynCalc} and \texttt{FeynArts} are distributed within the package installation but, just for completeness, their versions are 10.1.0 \textit{(stable version)} and 3.12, respectively. Note that the \texttt{FeynArts} version we use is the one directly provided with the \texttt{FeynCalc} installation and, therefore, it is already patched to \texttt{FeynCalc}.

Before running \tool, the user must specify the directory where the folder containing the package is located. You can do this by executing the following command:

\begin{mathematicaNotebook}
\begin{mmaCell}[]{Input}
  SetDirectory["path/to/mosca"];
\end{mmaCell}
\end{mathematicaNotebook}

\noindent
After setting the directory, load \tool \space by running the command:

\begin{mathematicaNotebook}
\begin{mmaCell}[moredefined=mosca]{Input}
  << mosca\`
\end{mmaCell}
\end{mathematicaNotebook}

\noindent
If everything is correct, you should see a GIF animation \footnote{You can disable the GIF animation by typing \texttt{\$LogoAnimation=False} before loading \tool.} on the screen drawing the \tool\ logo:

\begin{tcolorbox}[colback=black!2!white, sharp corners, left=-2mm, colframe=white!40!black, boxrule=0.3mm]
    \centering
    \includegraphics[width=0.35\linewidth]{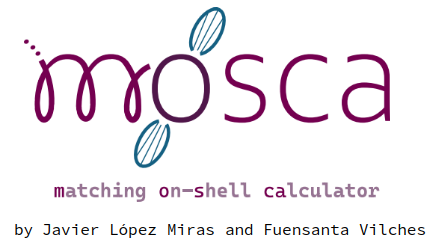}
\end{tcolorbox}

\noindent
You are now ready to start flying your \tool! \footnote{In Spanish, \textit{mosca} means ``fly'' 
\vcenteredinclude{12pt}{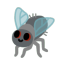}.}

\subsection{Setting up the Input}

\tool\ relies on the computation of Feynman diagrams. Hence, the first step consists on specifying a field content and an interaction Lagrangian. This is what we call a \textit{model}, and its structure follows that of \texttt{FeynArts}. The most straightforward way to create the models is to use the \mmaInlineCell{Input}{WriteFeynArtsOutput[]} function from \texttt{Feynrules}. When doing so, it is essential to have set the \mmaInlineCell{Input}{CouplingRename} option to \mmaInlineCell{Input}{False}. By default, this option is set to \mmaInlineCell{Input}{True}, which causes all amplitudes to be expressed in terms of unknown new variables that the package cannot process correctly. Additionally, to ensure proper handling of bilinear terms in the Feynman rules, such as 2-point functions, make sure that \mmaInlineCell{Input}{FR\$Loop=True} is also enabled. Once the models are generated, the \mmaInlineCell{Input}{FAPatch[]} function from \texttt{FeynCalc} can be used to patch them, ensuring compatibility with \texttt{FeynCalc}. If the version of \texttt{FeynArts} is already patched, the option \mmaInlineCell{Input}{PatchModelsOnly\(\,\,\)->\(\,\,\)True}  can be used to apply the patch exclusively to the model files.

By default, the 2-point function generation at tree-level via \texttt{Feynrules} requires a special modification of the models. These modifications have been fully automated within \tool. In particular, the list \texttt{M\$CouplingMatrices} from the model file (\texttt{.mod}), must be adapted. Each element of this list consists of a set of couplings corresponding to a certain Feynman rule or vertex. The set of couplings is arranged in turn in another list where each entry corresponds to a counter-term (CT) order. By default, the \texttt{FeynArts} model output generated from \texttt{Feynrules} does not include couplings for the 2-point function, even when \mmaInlineCell{Input}{FR\$Loop} is set to \mmaInlineCell{Input}{True} (this option adds the kinematic structures in the \texttt{.gen} file, but not the coupling vector counterpart in the \texttt{.mod}). However, there is a minor change that one can make in the \texttt{FeynArtsInterface.m} file of \texttt{Feynrules}: wherever the comment \mmaInlineCell{Input}{(*Remove the tree-level component for the vertices with two legs or less*)} appears, \linebreak one has to replace $>$ with $>=$ in the subsequent \mmaInlineCell{Input}{If[Length[#1]>2, \(\dots\)]} command. By doing this, the coupling vectors of 2-point vertices will be automatically written in the \texttt{.mod} file as 0th-order CTs. Otherwise, the user would have to provide these entries by hand.

Furthermore, the \mmaInlineCell{Input}{CreateTopologies[]} function  does not support a \mmaInlineCell{Input}{1->1} process at 0th-order CT. To address this, the \texttt{.mod} file must be modified by placing all coupling vectors corresponding to 2-point vertices under the entry associated with CTs of order 1 in the \texttt{M\$CouplingMatrices}. If we consider the following Lagrangian for a real scalar singlet $\phi$  with $Z_2$ symmetry:

\begin{equation}
    \mathcal{L}_{\phi} = -\frac{1}{2} \phi (\partial^2 + m^2) \phi + r_{D\phi} \partial^2 \phi \partial^2 \phi ~, 
    \label{eq: Lagrangian phi}
\end{equation}
the corresponding entry in the \texttt{M\$GenericCouplings} found in the \texttt{.gen} file should be given by:

\txtbox{81}{Extract of the 2-pt interaction kinematics in the \texttt{.gen} file}{
\tab (* SS *)

\tab AnalyticalCoupling[s1 S[j1, mom1], s2 S[j2, mom2] ] == G[+1][s1 S[j1], s2 S[j2]].\{ 

\tab\tab 1, 

\tab\tab FAScalarProduct[mom1,mom2], 

\tab\tab FAScalarProduct[mom1,mom1]FAScalarProduct[mom2,mom2] \}
}
\noindent
On the other hand, the associated entry in the \texttt{M\$CouplingMatrices} list in the \texttt{.mod} file should be:

\txtbox{56}{Extract of the 2-pt interaction coupling in the \texttt{.mod} file}{
\tab C[ S[1] , S[1] ] == \{

\tab\tab \{0, (-I)*m\textasciicircum 2\}, 

\tab\tab \{0, -I\}, 

\tab\tab \{0, (2*I)*rD\textbackslash[Phi]\} \}
}

\noindent
Here, the couplings for each 2-point interaction are placed in the second position of the list, ensuring they are treated as 1st-order CTs. The \mmaInlineCell[moredefined=model]{Input}{Patch2ptFunction[model]} function in \tool \space automates this procedure, taking the 2-point coupling vectors from a given \texttt{model.mod} file and applying the necessary modification. This function can be executed alongside other patch functions using \mmaInlineCell[moredefined=model]{Input}{Patch2mosca[model]}. The \mmaInlineCell[moredefined=model]{Input}{Patch2mosca[model]} command also includes the functionalities of \mmaInlineCell[moredefined=model]{Input}{PatchMasses[model]}, which assigns a unique symbolic internal mass to every field (massless or not) to ensure proper differentiation within the calculations, together with the \texttt{FeynCalc} \mmaInlineCell[moredefined=model]{Input}{FAPatch[]}.

To ensure the correct operation of all functions, strict naming conventions must be followed. The \texttt{.gen} and \texttt{.mod} files must share the same base name, omitting the file extension when referring to them. To simplify the access, it is convenient to define a variable that specifies the path to these files. For instance, the command 
\begin{mathematicaNotebook}
\begin{mmaCell}[moredefined={path}]{Input}
  model = FileNameJoin[\{path, "Models", "model1", "model1"\}];
\end{mmaCell}
\end{mathematicaNotebook}
\noindent
can be used to reference the files \texttt{model1.gen} and \texttt{model1.mod}, which are stored in the \texttt{model1} folder within the \texttt{Models} folder directory specified by \texttt{path}. From now on, when we refer to a model in a \mathematica\ context (\textit{e.g.}, as arguments of functions) we mean exactly this: the string containing the path where the model \texttt{.gen} and \texttt{.mod} files are found.

Once all these considerations are implemented, the final step is to define the couplings present in the models. This is achieved by using the following function:

\begin{functiondef}
\mmaInlineCell[moredefined={coefficient,model,X}]{Input}{DefineWC[coefficient, model, EFTOrder\(\,\,\)->\(\,\,\)X, ComplexParameter\(\,\,\)->\(\,\,\)True/False]}
\end{functiondef}

\noindent
which takes two inputs, namely, the name of a coefficient and the model it is associated with, and two optional arguments: the corresponding EFT order and whether the coefficient is complex. Hereafter, the EFT order $X$ of the dimensionless coupling $c$, appearing in the Lagrangian as $\frac{c}{\Lambda^n} \mathcal{O}_i$, is understood to be $X=n+4$, unless it is the WC of a renormalizable operator in which case $X=4$ independently of the mass dimension of $O_i$. The function can also handle a list of coefficients, provided they all share the same \mmaInlineCell{Input}{EFTOrder} and \mmaInlineCell{Input}{ComplexParameter} settings. If these options are omitted, the function defaults to \mmaInlineCell{Input}{EFTOrder\(\,\,\)->\(\,\,\)4} and \mmaInlineCell{Input}{ComplexParameter\(\,\,\)->\(\,\,\)False}. A more convenient way to define all the coefficients in a model is by using a structured list. For instance, a real renormalizable coupling, $c_1$, and a complex dimension-6 coupling, $c_2$, can be defined via the following list:

{\small
\begin{tcolorbox}[breakable,colback=white,middle=0mm,boxsep=2mm,colframe=white!30!black,colbacktitle=white!50!black,fonttitle=\bfseries,title=\normalsize{Example of parameter file to define the WCs of a model}]
\setlength\linenumbersep{-.1cm} 
\setcounter{linenumber}{1}
\begin{internallinenumbers}
\begin{verbatim}
    paramList = {
      c1 == {
        ComplexParameter -> False,
        EFTOrder -> 4 },
        
      c2 == {
        ComplexParameter -> True,
        EFTOrder -> 6 },
      ...
\end{verbatim}
\setcounter{linenumber}{20}
\begin{verbatim}
    };
\end{verbatim}
\end{internallinenumbers}
\end{tcolorbox}
}

\noindent
This format is similar to the \texttt{M\$Parameters} list from the \texttt{.fr} file used in \texttt{FeynRules} for model creation, which is very convenient in order to reuse its structure (note that any option other than \mmaInlineCell{Input}{ComplexParameter} and \mmaInlineCell{Input}{EFTOrder} is simply ignored). The name of the variable at which the list is assigned is unimportant (one could even write the list without any assignment), but this must be the only element present in the file. Then, the function 

\begin{functiondef}
    \mmaInlineCell[moredefined={model,paramListPath}]{Input}{DefineAllWCs[model, paramListPath]}
\end{functiondef}

\noindent
automates the process by taking the model and the path to the file containing \texttt{paramList}. It then extracts the necessary information for all coefficients. If any options are omitted, the function applies the default settings. Finally, by executing \texttt{\textbf{WClist[model]}} one can access the full list of defined WCs for the specified model. It is important to notice that WCs are interpreted in \tool\ as objects with head \mmaInlineCell{Input}{WC[]} and three arguments, namely, the symbol of the coupling $c$, its EFT order $X$ and the model it belongs to. We keep the models so that users can freely have equal coupling symbols in different Lagrangians, facilitating the creation of models and avoiding possible mix-ups. In the model creation, however, one should write the couplings simply with their symbol name $c$. The next section will cover how \tool\ converts symbols, $c$, into appropriate objects \mmaInlineCell[moredefined={c,X,model}]{Input}{WC[c,X,model]} and how to format and present the results in a readable manner.

\subsection{Customizing the Output}

We first introduce various options for presenting the outputs. The primary function for this purpose is

\begin{functiondef}
    \mmaInlineCell[moredefined=model]{Input}{FormatWC[model, ModelName\(\,\,\)->\(\,\,\)"", VisibleModel\(\,\,\)->\(\,\,\)True/False,}
    \mmaInlineCell{Input}{VisibleOrder\(\,\,\)->\(\,\,\)True/False]}
\end{functiondef}

\noindent
which allows setting a specific format for all the WCs within a model. The function admits three optional inputs: the name of the model, provided as a string tagging the model of the coefficients; and two boolean options, \mmaInlineCell{Input}{VisibleModel} and \mmaInlineCell{Input}{VisibleOrder}. The first one controls whether the chosen model name is displayed as a subscript, while the second determines whether the EFT order of the coefficients is shown inside a parenthesis as a superscript. Both of these options are set to \mmaInlineCell{Input}{True} by default. However, \mmaInlineCell{Input}{VisibleModel} will not display any subscript if \mmaInlineCell{Input}{ModelName} is not specified, since the default value for this input is an empty string.

As an example, let us print \mmaInlineCell[moredefined=modelPhi]{Input}{WClist[modelPhi]} in a \mathematica\ notebook, where \mmaInlineCell[moredefined=modelPhi]{Input}{modelPhi} is the Lagrangian in Eq. \eqref{eq: Lagrangian phi}. This list contains all the WCs of the Lagrangian, after having defined them via the \mmaInlineCell{Input}{DefineWC[]} commands. If we do not run the \mmaInlineCell{Input}{FormatWC[]} function, we obtain:

\begin{mathematicaNotebook}
\begin{mmaCell}[moredefined={modelPhi}]{Input}
  WClist[modelPhi]
\end{mmaCell}
\begin{mmaCell}{Output}
  \{ WC[m, 4, "path-to-mosca/Models/modelPhi/modelPhi"],
     WC[rD\(\phi\), 6, "path-to-mosca/Models/modelPhi/modelPhi"] \}
     
\end{mmaCell}
\end{mathematicaNotebook}

\noindent
To properly format the WCs output, we use:

\begin{mathematicaNotebook}
\begin{mmaCell}[moredefined={modelPhi}]{Input}
  FormatWC[modelPhi, ModelName->"model\(\phi\)"];
\end{mmaCell}
\end{mathematicaNotebook}

\noindent
which now produces:

\begin{mathematicaNotebook}
\begin{mmaCell}[moredefined={modelPhi}]{Input}
  WClist[modelPhi]
\end{mmaCell}
\begin{mmaCell}{Output}
  \{ \mmaSubSup{m}{model\(\phi\)}{(4)}, \mmaSubSup{rD\(\phi\)}{model\(\phi\)}{(6)} \}
\end{mmaCell}
\end{mathematicaNotebook}

\noindent
If we prefer to hide the EFT order in the output, we can set the \mmaInlineCell{Input}{VisibleOrder} option to \mmaInlineCell{Input}{False}:

\begin{mathematicaNotebook}
\begin{mmaCell}[moredefined={modelPhi}]{Input}
  FormatWC[modelPhi, ModelName->"model\(\phi\)", VisibleOrder->False];
\end{mmaCell}
\end{mathematicaNotebook}

\noindent
Now, the coefficients are shown in the following way:

\begin{mathematicaNotebook}
\begin{mmaCell}{Output}
  \{ \mmaSub{m}{model\(\phi\)}, \mmaSub{rD\(\phi\)}{model\(\phi\)} \}
\end{mmaCell}
\end{mathematicaNotebook}

\noindent
Similarly, to omit the model subscript, we can set \mmaInlineCell{Input}{VisibleModel\(\,\,\)->\(\,\,\)False}. In general, for the rest of the examples, we will assume both options are set to \mmaInlineCell{Input}{False}, unless stated otherwise. One can always check the true form of the WCs with an \mmaInlineCell{Input}{InputForm[]} command.

\subsubsection{Treatment of coefficients}

The function \mmaInlineCell[moredefined=model]{Input}{RenameWC[model]} generates a substitution rule to rewrite the WCs of an expression into the proper format required for processing by \tool, this is, objects with head \mmaInlineCell{Input}{WC[]}. For instance, suppose that a WC $c$ has been defined, \textit{e.g.} using \mmaInlineCell[moredefined={modelC,c,X}]{Input}{DefineWC[c, modelC, EFTOrder\(\,\,\)->\(\,\,\)X]}. If we then write an expression containing $c$, say \mmaInlineCell[moredefined=exp]{Input}{exp}, we can run:

\begin{mathematicaNotebook}
\begin{mmaCell}[moredefined={exp, RenameWC, modelC}]{Input}
  exp /. RenameWC[modelC] 
\end{mmaCell}
\end{mathematicaNotebook}

\noindent
to replace every occurrence of $c$ in \mmaInlineCell[moredefined=exp]{Input}{exp} with the correctly formatted \mmaInlineCell[moredefined={c,X,modelC}]{Input}{WC[c, X, modelC]}.

Additionally, the function \mmaInlineCell[moredefined=exp]{Input}{ExplicitEFTOrder[exp]} multiplies each WC in \mmaInlineCell[moredefined=exp]{Input}{exp} by the corresponding powers of \mmaInlineCell{Input}{inv\(\Lambda\)}, based on its dimension. Applying both transformations in sequence to \mmaInlineCell[moredefined={exp,A,B,c}]{Input}{exp = A\(\,\)c\(\,\,\)+\(\,\,\)B\(\,\)c^2} produces:

\begin{mathematicaNotebook}
\begin{mmaCell}[moredefined={exp, RenameWC, modelC}]{Input}
  exp /. RenameWC[modelC] //ExplicitEFTOrder
\end{mmaCell}
\begin{mmaCell}[moredefined={exp, RenameWC, modelC}]{Output}
  A \mmaSup{inv\(\Lambda\)}{X-4} WC[c, X, modelC] + B \mmaSup{(\mmaSup{inv\(\Lambda\)}{X-4})}{2} \mmaSup{WC[c, X, modelC]}{2}
\end{mmaCell}
\end{mathematicaNotebook}

\noindent
Here, we have left the output unformatted to clearly show that the WCs have been correctly replaced. This last function is described in detail in Section~\ref{sect: Usefil Functions}.

Another useful function is \mmaInlineCell{Input}{TrueEFTOrder[]}, which determines the EFT order associated with a given WC. While this function may initially seem unnecessary, its purpose becomes clear when understanding how \tool \space systematically solves the system of equations to find the redefinition of the WCs. As a brief preview, \tool \space performs the reduction process by solving for each coefficient order-by-order in the EFT expansion. Specifically, instead of solving directly for $c^{(X)}_{\mathrm{modelC}}$, we express the coefficient as a sum of contributions at different orders:
\begin{equation*}
    \frac{1}{\Lambda^{X-4}} c^{(X)}_{\mathrm{modelC}} \to \frac{1}{\Lambda^{X-4}} \left( c^{(X)}_{\mathrm{modelC}} + \frac{1}{\Lambda} c^{(X+1)}_{\mathrm{modelC}} + \frac{1}{\Lambda^{2}} c^{(X+2)}_{\mathrm{modelC}} + \dots + \frac{1}{\Lambda^{N-X-1}} c^{(N-1)}_{\mathrm{modelC}} + \frac{1}{\Lambda^{N-X}} c^{(N)}_{\mathrm{modelC}} \right)~, 
\end{equation*}
\noindent
where each $c^{(n)}_{\mathrm{modelC}}$ (with $n \in [X,N]$, being $N$ the order at which we truncate the EFT expansion) represents a distinct contribution to the redefinition of the WC $c$ at dimension $n$. However, the EFT order of a WC is determined solely by the operator it accompanies and is independent of $n$. Then, given a specific coefficient $c^{(n)}_{\mathrm{modelC}}$, we can retrieve its true EFT order by using:

\begin{mathematicaNotebook}
\begin{mmaCell}[moredefined={TrueEFTOrder, WC, c, n, modelC}]{Input}
  TrueEFTOrder[WC[c, n, modelC]]
\end{mmaCell}
\begin{mmaCell}[moredefined={exp, RenameWC, modelC}]{Output }
  X
\end{mmaCell}
\end{mathematicaNotebook}

\section{A Guide to \tool: Objects and Functions}
\label{sect: guide}
\subsection{2-point Functions Calculation}

A key feature that distinguishes on-shell matching is the need of computing what in the following will be referred to as 2-point attributes, namely, physical masses, wavefunction factors and explicit forms of Feynman propagators in momentum space. The precise expressions of the masses are needed for the correct generation of on-shell momenta, while the wavefunction factors appear in the computation of scattering amplitudes according to the LSZ formula. Meanwhile, propagators should account for the 2-point interactions in internal propagator lines of Feynman diagrams. In a given model, all operators involving two fields contribute to these functions, which means that these attributes will generally differ for each case. Hence, it is essential to adopt a consistent approach to incorporate these effects.

Propagators are obtained (up to factors of $i$) as the inverse of the 2-point functions, which we have access to by computing the amplitude $\phi \to \phi$ at tree-level and CT order 1 in \texttt{FeynArts}, with $\phi$ a generic field. While the propagator for scalar fields  can be computed straightforwardly by inverting the 2-point function, the process becomes more complex for vector and fermion fields due to the tensor nature of these propagators. 

First, in the case of a scalar, if we add arbitrary 2-point (local) operators we will have a 2-point Green's function of the form $\Gamma_S=p^2 - R(p^2)$, where $R$ is a polynomial (\textit{e.g.}, in a massive free scalar theory we would have $R=m^2$). Therefore, the scalar propagator is simply
\begin{equation}
    D(p) = \frac{i}{p^2 - R(p^2)}~.
    \label{eq:scalar propagator}
\end{equation}
Now, let us move on to vectors. Given a generic 2-point function $\Gamma_V^{\mu\nu}$ of a vector field, the corresponding propagator $\Pi_{\mu\nu}$ is determined by imposing $ \Gamma_V^{\mu\rho}\Pi_{\rho\nu}=i\delta^\mu_\nu$. In general, $\Gamma_V^{\mu\nu}$ and $\Pi_{\mu\nu}$ can depend on all possible Lorentz structures:
\begin{equation*}
    \Gamma_V^{\mu\nu}=A_1 g^{\mu\nu} + A_2 p^\mu p^\nu ~, \qquad
    \Pi_{\mu\nu}=B_1 g_{\mu\nu} + B_2 p_\mu p_\nu ~,
\end{equation*}
where all coefficients can possibly depend on $p^2$. $A_1$ and $A_2$ can be separated into two distinct components: a canonical part coming from the kinetic and gauge-fixing terms, and a second part resulting from the contribution of other (effective) operators. We will denote the non-canonical parts as $R_1$ and $R_2$, so we have that
\begin{equation*}
    \Gamma_V^{\mu\nu}=\left(-p^2+R_1 \right) g^{\mu\nu} + \left(1-\frac{1}{\xi}+R_2 \right) p^\mu p^\nu ~.
\end{equation*}
Doing now
\begin{equation*}
    i \delta^\mu_\nu=\Gamma_V^{\mu\rho}\Pi_{\rho\nu} = B_1  \left(-p^2+R_1 \right) \delta^\mu_\nu + \left[ B_1  \left(1-\frac{1}{\xi}+R_2 \right) + B_2 \left(R_1 + p^2 R_2 -\frac{p^2}{\xi} \right) \right] p^\mu p_ \nu ~,
\end{equation*}
we obtain that 
\begin{equation*}
    \Pi_{\mu\nu}=- \frac{ i g_{\mu\nu}}{p^2-R_1}- p_\mu p_\nu \frac{i (1-\xi+R_2 \xi)(\xi R_1 + p^2 \xi R_2 - p^2)^{-1}}{p^2-R_1} ~.
\end{equation*}
By setting $\xi=0$, we find that the non-canonical contributions to the numerator vanish, and so does every dependence on $R_2$ in $\Pi_{\mu\nu}$. Then, in the Landau gauge, the modification in the vector propagators only manifests as a modification in the denominator given by the component of the 2-point function proportional to $g^{\mu\nu}$:
\begin{equation}
    \Pi_{\mu\nu}(p) = \frac{-i\left(g_{\mu\nu} - \frac{p_{\mu}p_{\nu}}{p^2}\right)}{p^2-R_1(p^2)}~.
    \label{eq:vector propagator}
\end{equation}
This makes the Landau gauge a very convenient choice for simplification of amplitudes.

For fermion fields, the general form of the 2-point function can be expressed as
\begin{equation*}
    \Gamma_F =  \slashed{p} A + B ~,
\end{equation*}
where $A,B$ are $4\times4$ Dirac structures functions of the squared-momentum. Due to the absence of free Lorentz indices, without loss of generality $A$ and $B$ must be linear combinations of the chirality projectors $P_R$ and $P_L$. Therefore, we can write $$A = R_1 P_R +\widetilde{R}_{1} P_L, \qquad B = R_2 P_R +\widetilde{R}_{2} P_L\,,$$ where $R_i,\widetilde{R}_{i}$ are polynomials of $p^2$. Inverting this expression yields the fermion propagator:
\begin{equation}
    S(p) = \frac{i}{ \slashed{p} (R_1 P_R + \widetilde{R}_{1}P_L) + (R_2 P_L+\widetilde{R}_{2}P_R)} = 
     \frac{ (\slashed{p} R_1 - R_2) P_R + (\slashed{p} \widetilde{R}_{1} - \widetilde{R}_{2} ) P_L}{R_1 \widetilde{R}_1 p^2 - R_2 \widetilde{R}_{2} }~.
    \label{eq:fermion propagator}
\end{equation}
Here, both the numerator and denominator of the propagator are influenced by the non-canonical contributions to $R_i$ and $\widetilde{R}_{i}$.

Once we know how to determine the explicit forms for propagators from the 2-point Green's functions, physical masses are obtained as the poles of such expressions while the wavefunction $Z$ factors are the residues at these poles, as it was explained in Section \ref{sect: intro}.

In \tool \space we incorporated all this theoretical framework to automatically find the physical mass, the wave function renormalization factor and the propagator for each field from the 2-point Feynman rules of the model. By running

\begin{functiondef}
\mmaInlineCell[moredefined={field,model}]{Input}{PropagatorAttributes[field, model, EFTOrder, Output\(\,\,\)->\(\,\,\)True/False]}
\end{functiondef}
\noindent
one can retrieve all the attributes associated with the 2-point function of a given field in the specified \mmaInlineCell[moredefined=model]{Input}{model}, up to the dimension determined by \mmaInlineCell{Input}{EFTOrder}. The \mmaInlineCell{Input}{Output} option, enabled by default (\mmaInlineCell{Input}{True}), formats and displays the result. For the Lagrangian of equation \eqref{eq: Lagrangian phi} we would have 

\begin{mathematicaNotebook}
\begin{mmaCell}[moredefined={modelPhi}]{Input}
  \{mass,Zwf,prop\} = PropagatorAttributes[S[1], modelPhi, 6]
\end{mmaCell}
\begin{mmaCell}{Print}
  \textbf{Propagator Attributes of S[1]:}
  \{ \mmaSup{\mmaSub{m}{phys}}{2} \(\to\) \mmaSup{m}{2}\(\,\,\)-\(\,\,\)2\(\,\,\)\mmaSup{inv\(\pmb{\Lambda}\)}{2}\mmaSup{m}{4}rD\(\phi\), Z \(\to\) 1\(\,\,\)-\(\,\,\)4\(\,\,\)\mmaSup{inv\(\pmb{\Lambda}\)}{2}\mmaSup{m}{2}rD\(\phi\), Prop \(\to\) \mmaFrac{1}{\mmaSup{p}{2}-\mmaSup{m}{2}}\(\,\,\)-\(\,\,\)\mmaFrac{2\(\,\,\)\mmaSup{p}{4}\mmaSup{inv\(\pmb{\Lambda}\)}{2}rD\(\phi\)}{\mmaSup{(\mmaSup{p}{2}-\mmaSup{m}{2})}{2}} \}

\end{mmaCell}
\begin{mmaCell}{Output}
  \{ \mmaSup{m}{2}\(\,\,\)-\(\,\,\)2\(\,\,\)\mmaSup{inv\(\pmb{\Lambda}\)}{2}\mmaSup{m}{4}rD\(\phi\), 1\(\,\,\)-\(\,\,\)4\(\,\,\)\mmaSup{inv\(\pmb{\Lambda}\)}{2}\mmaSup{m}{2}rD\(\phi\), Prop\$23001 \}  
\end{mmaCell}
\end{mathematicaNotebook}

\noindent
The function returns the physical mass, the wavefunction renormalization and the propagator (notice that the list appearing before the \textit{Out} cell is simply a \textit{Print} display). The latter is a \mathematica\ \textit{pure function} whose explicit form can be recovered upon evaluation at the momentum and the momentum squared \footnote{In practice it would only be necessary to specify the momentum. Adding the squared momentum as a second variable is only a matter of convenience to correctly deal with \texttt{SP[mom,mom]} structures instead of \texttt{mom\textasciicircum 2}. Even so, it is possible to call also the propagator with just one variable, say $p$, which will be understood internally as a call with $p$ and $p^2$. Furthermore, a non-squared momentum $p$ can only appear in fermion propagators, as we will see.}:

\begin{mathematicaNotebook}
\begin{mmaCell}[moredefined={prop}]{Input}
  prop[p, p2]
\end{mmaCell}
\begin{mmaCell}{Output}
  \mmaFrac{1}{p2-\mmaSup{m}{2}}\(\,\,\)-\(\,\,\)\mmaFrac{2\(\,\,\)\mmaSup{p2}{2}\mmaSup{inv\(\pmb{\Lambda}\)}{2}rD\(\phi\)}{\mmaSup{(p2-\mmaSup{m}{2})}{2}}
\end{mmaCell}
\end{mathematicaNotebook}

\noindent
Not only does \mmaInlineCell{Input}{PropagatorAttributes[]} compute, show and return the 2-point attributes, but it also updates the reserved keywords \mmaInlineCell[moredefined={field,model}]{Input}{mphys2[field,model]}, \mmaInlineCell[moredefined={field,model}]{Input}{Z[field,model]} and \mmaInlineCell[moredefined={field,model}]{Input}{Prop[field,model]}, which can be called to access to the different parameters of a field in a specific model. 

It should be emphasized what we mean by ``propagator'' in this context. Actually, the object \mmaInlineCell[moredefined={field,model}]{Input}{Prop[field,model]} is thought to be the replacement for the generic
\mmaInlineCell{Input}{FeynAmpDenonimator}s 
\texttt{FeynCalc} structures in amplitudes. Thus, for scalars and vectors (in the Landau gauge), where only the denominator changes, \mmaInlineCell{Input}{Prop[]} is simply the corresponding denominator. However, for fermions, according to \eqref{eq:fermion propagator} we need to modify the numerator too. Hence, we identify every $\slashed{p}\pm m$ (coming from a fermion propagator) and replace it---keeping its position in the fermionic chain--- by the propagator dictated by Eq.~\eqref{eq:fermion propagator} (up to the factor of $i$). The precise expressions for the \mmaInlineCell{Input}{Prop} objects for scalar, spinor and vector fields can be found in Table \ref{tab:Prop expressions}.

\begin{table}[ht]
    \centering
    \begin{tabularx}{0.8\linewidth}{YYc}
        \toprule
      \textbf{Scalar} & \textbf{Vector} & \textbf{Fermion} \\ \midrule
       $\dfrac{1}{p^2-R}$ & $\dfrac{1}{p^2-R_1}$  & $\dfrac{ (\slashed{p} R_1 - R_2) P_R + (\slashed{p} \widetilde{R}_{1} - \widetilde{R}_{2} ) P_L}{R_1 \widetilde{R}_1 p^2 - R_2 \widetilde{R}_{2} }$ \\[1ex]
       \bottomrule
    \end{tabularx}
    \caption{Expressions for \textbf{\texttt{Prop}} in \tool. These yield the correct propagators for scalars, vectors and fermions upon multiplication by $i$, $-i (g_{\mu\nu}-p_\mu p_\nu /p^2)$ and $i$, respectively. Refer to the beginning of this section for further details in the definition of $R,R_i$ and $\widetilde{R}_i$ in each case.}
    \label{tab:Prop expressions}
\end{table}

If the user wants to compute all these attributes for every field within a model at the same time, then it is enough to execute

\begin{functiondef}
\mmaInlineCell[moredefined=model]{Input}{AllPropagatorAttributes[model, EFTOrder, ExcludeParticles\(\,\,\)->\(\,\,\)\{\},}
\mmaInlineCell{Input}{Output\(\,\,\)->\(\,\,\)True/False]}
\end{functiondef}

\noindent
The \mmaInlineCell{Input}{ExcludeParticles} option allows to omit any field present in the model for the computation. For example, we extend the Lagrangian of equation \eqref{eq: Lagrangian phi} to

\begin{equation}
    \mathcal{L}_{full} = \mathcal{L}_\phi - \frac{1}{4} F_{\mu\nu} F^{\mu\nu} + \overline{\psi} (i \slashed{D} - m_\psi) \psi - \frac{1}{2} r_{2F} \partial_\mu F^{\mu\nu} \partial^\rho F_{\rho\nu} + r_{{\psi}D} \overline\psi i \{D^2, \slashed{D}\} \psi + \mathcal{L}_{int}~,
    \label{eq: Lagrangian full}
\end{equation}
where $\mathcal{L}_{int}$ will be defined later, since it contains operators with more than two fields which are not relevant now. Here, the fermion $\psi$ is coupled to a $U(1)$ gauge symmetry and the corresponding field-strength tensor is given by $F_{\mu\nu}$. Thus, in the \texttt{FeynArts} model we have a scalar \mmaInlineCell{Input}{S[1]}, a fermion \mmaInlineCell{Input}{F[1]} and a vector \mmaInlineCell{Input}{V[1]}.  Naming \mmaInlineCell[moredefined=modelFull]{Input}{modelFull} the model associated with this Lagrangian, we would obtain the result by simply running

\begin{mathematicaNotebook}
\begin{mmaCell}[moredefined={modelFull}]{Input}
  AllPropagatorAttributes[modelFull, 6];
\end{mmaCell}
\begin{mmaCell}{Print}
  \textbf{Propagator Attributes of F[1]:}
  \{ \mmaSubSup{m}{phys}{2} \(\to\) \mmaSup{m\(\psi\)}{2}\(\,\,\)+\(\,\,\)2\mmaSup{inv\(\pmb{\Lambda}\)}{2}\mmaSup{m\(\psi\)}{4}r\(\psi\)D, Z \(\to\) 1\(\,\,\)+\(\,\,\)3\mmaSup{inv\(\pmb{\Lambda}\)}{2}\mmaSup{m\(\psi\)}{2}r\(\psi\)D, 
  Prop \(\to\) \mmaFrac{(\(\gamma\) \(\cdot\) p)+m\(\psi\)}{\mmaSup{p}{2}-\mmaSup{m\(\psi\)}{2}}\(\,\,\)+\(\,\,\)\mmaFrac{\mmaSup{inv\(\pmb{\Lambda}\)}{2}(\mmaSup{p}{4}(\(\gamma\) \(\cdot\) p)\(\,\,\)+\(\,\,\)2\mmaSup{p}{4}m\(\psi\)\(\,\,\)+\(\,\,\)\mmaSup{p}{2}(\(\gamma\) \(\cdot\) p)\mmaSup{m\(\psi\)}{2})r\(\psi\)D}{\mmaSup{(\mmaSup{p}{2}-\mmaSup{m\(\psi\)}{2})}{2}} \} \\
  
  \textbf{Propagator Attributes of S[1]:}
  \{ \mmaSubSup{m}{phys}{2} \(\to\) \mmaSup{m}{2}\(\,\,\)-\(\,\,\)2\(\,\,\)\mmaSup{inv\(\pmb{\Lambda}\)}{2}\mmaSup{m}{4}rD\(\phi\), Z \(\to\) 1\(\,\,\)-\(\,\,\)4\(\,\,\)\mmaSup{inv\(\pmb{\Lambda}\)}{2}\mmaSup{m}{2}rD\(\phi\), Prop \(\to\) \mmaFrac{1}{\mmaSup{p}{2}-\(\,\,\)\mmaSup{m}{2}}\(\,\,\)-\(\,\,\)\mmaFrac{2\(\,\,\)\mmaSup{p}{4}\mmaSup{inv\(\pmb{\Lambda}\)}{2}rD\(\phi\)}{\mmaSup{(\mmaSup{p}{2}-\mmaSup{m}{2})}{2}} \}\\

  \textbf{Propagator Attributes of V[1]:}
  \{ \mmaSubSup{m}{phys}{2} \(\to\) 0, Z \(\to\) 1, Prop \(\to\) \mmaFrac{1}{\mmaSup{p}{2}}\(\,\,\)-\(\,\,\)\mmaSup{inv\(\pmb{\Lambda}\)}{2}r2F \} 

\end{mmaCell}
\end{mathematicaNotebook}

\noindent

The \mmaInlineCell{Input}{mphys2[]} object gives the expression for the physical mass in terms of the bare mass and other coefficients of the Lagrangian. It is also important to have the inverse relation, namely, the bare mass in terms of the physical mass (up to a given EFT order). The function that handles this is
\begin{functiondef}
    \mmaInlineCell[moredefined={field,model}]{Input}{Bare2PhysMass[field,\(\,\,\)EFTOrder,\(\,\,\)model]}
\end{functiondef}
\noindent
It accepts either a single field or a list of fields as input. The redefinition (returned as a replacement rule) is given in terms of \mmaInlineCell[moredefined=field]{Input}{MPhysSymbol[field]}, which represents the symbolic physical mass assigned internally to each field during the different computations. This function applied to the scalar field in the model yields the following result:

\begin{mathematicaNotebook}
\begin{mmaCell}[moredefined=modelFull]{Input}
  Bare2PhysMass[S[1], 6, modelFull]
\end{mmaCell}
\begin{mmaCell}[moredefined=AllMassReduction]{Output}
  m \(\to\) MPhysSymbol[S[1]]\(\,\,\)+\(\,\,\)\mmaSup{ MPhysSymbol[S[1]]}{3} rD\(\phi\) 
\end{mmaCell}
\end{mathematicaNotebook}

On the other hand, the functions \mmaInlineCell[moredefined={expression,model}]{Input}{ReplaceBareMass[expression,\(\,\,\)EFTOrder,\(\,\,\)model]} and \linebreak \mmaInlineCell[moredefined={expression,model}]{Input}{ReplacePhysMass[expression,\(\,\,\)EFTOrder,\(\,\,\)model]} are responsible for rewriting an expression by translating bare masses of a given model in terms of physical masses, and \textit{vice versa}.
They are specifically designed to handle expressions that depend on \mmaInlineCell{Input}{MPhysSymbol[]}, which is useful since, in general, the partial results arising from \tool's matching functions are expressed using this symbolic physical mass. Both \mmaInlineCell{Input}{ReplacePhysMass[]} and \mmaInlineCell{Input}{ReplaceBareMass[]} also automatically expand the result up to the dimension specified by \mmaInlineCell{Input}{EFTOrder}. For example, given an expression \mmaInlineCell[moredefined={exp,a,b}]{Input}{exp}, 
applying \mmaInlineCell{Input}{ReplacePhysMass[]} produces the following result: 

\begin{mathematicaNotebook}
\begin{mmaCell}[moredefined={exp,modelFull,ReplacePhysMass}]{Input}
  exp\(\,\,\)=\(\,\,\)a\(\,\,\)MPhysSymbol[S[1]]\(\,\,\)+\(\,\)b MPhysSymbol[S[1]]^2;
  ReplacePhysMass[exp, 6, modelFull]
  
\end{mmaCell}
\begin{mmaCell}[moredefined=AllMassReduction]{Output}
  a\(\,\,\)(m\(\,\,\)-\(\,\,\)\mmaSup{m}{3}\(\,\,\)rD\(\phi\))\(\,\,\)+\(\,\,\)b\(\,\,\)(\mmaSup{m}{2}-\(\,\,\)2\(\,\,\)\mmaSup{m}{4}\(\,\,\)rD\(\phi\))
\end{mmaCell}
\end{mathematicaNotebook}

\subsection{Amplitude Computation}

Once the tools regarding 2-point amplitudes have been discussed we move on to S-matrix elements of general processes. The amplitudes can be computed using the following command:

\begin{functiondef}
\mmaInlineCell[moredefined={process,model}]{Input}{AmplitudeComputation[process,\(\,\,\)EFTOrder,\(\,\,\)model,\(\,\,\)ExcludeParticles\(\,\,\)->\(\,\,\)\{\},\(\,\,\)}
\mmaInlineCell{Input}{SeparateCrossings\(\,\,\)->\(\,\,\)True/False]}
\end{functiondef}

\noindent  
This function calculates the physical amplitude for the specified process, considering all particles incoming, up to the given \mmaInlineCell{Input}{EFTOrder} within the chosen model. The option \mmaInlineCell{Input}{ExcludeParticles} allows for the exclusion of specific particles by specifying the corresponding fields. By default, no particles are excluded, as \mmaInlineCell{Input}{ExcludeParticles} is set to an empty list. Additionally, the \mmaInlineCell{Input}{SeparateCrossings} option is set to \mmaInlineCell{Input}{False} by default, meaning that the amplitude is returned as a single expression. If \mmaInlineCell{Input}{SeparateCrossings} is set to \mmaInlineCell{Input}{True}, the function will return the result separating each diagram and all possible permutations of its external legs (\textit{crossings}). We will cover this in detail in the next subsection.

Before presenting some examples, we must introduce the interaction term \( \mathcal{L}_{int} \) from the Lagrangian in equation \eqref{eq: Lagrangian full}, where

\begin{equation}
\mathcal{L}_{int} = a_{{\psi}F} F_{\mu \nu} \overline{\psi} \sigma^{\mu\nu} \psi + a_{\phi} \phi ^6 + a_\psi \overline{\psi} \gamma^\mu \psi \overline{\psi} \gamma_\mu \psi +r_{\phi D} \phi^3 \partial^2\phi + r_{DF\psi} D^\mu F_{\mu \nu} \overline{\psi} \gamma^\nu \psi~~.
\label{eq: Lagrangian interaction}
\end{equation}
Additionally, we define \mmaInlineCell[moredefined={modelPhys}]{Input}{modelPhys} as the model containing the Lagrangians \eqref{eq: Lagrangian full} and \eqref{eq: Lagrangian interaction} with all redundant operators (those with WCs $r_i$) set to zero.

Given then the following input:

\begin{mathematicaNotebook}
\begin{mmaCell}[moredefined={modelPhys}]{Input}
  AmplitudeComputation[\{F[1], -F[1], V[1]\}, 6, modelPhys] // TraditionalForm
\end{mmaCell}
\end{mathematicaNotebook}

\noindent
The function generates the amplitude \footnote{In the remaining examples of this subsection, for brevity, we will suppose that $a_{{\psi}F}=0$.}:

\begin{mathematicaNotebook}
\begin{mmaCell}[form=TraditionalForm]{Output}
\end{mmaCell}
\[
g\left( -(\varphi (-\overline{P2})) \cdot (\overline{\gamma}\cdot\overline{\varepsilon}(P3))\cdot\overline{\gamma}^6\cdot(\varphi (\overline{P1}))\right)
- g\left((\varphi (-\overline{P2})) \cdot (\overline{\gamma}\cdot\overline{\varepsilon}(P3))\cdot\overline{\gamma}^7\cdot(\varphi (\overline{P1}))\right)
\]
\end{mathematicaNotebook}

\noindent
If we set \mmaInlineCell{Input}{SeparateCrossings->True}, the result would rather be:

\begin{mathematicaNotebook}
\begin{mmaCell}[form=TraditionalForm]{Output}
\end{mmaCell}
\[
\begin{pmatrix}
\{ g\left( -(\varphi (-\overline{P2})) \cdot (\overline{\gamma}\cdot\overline{\varepsilon}(P3))\cdot\overline{\gamma}^6\cdot(\varphi (\overline{P1}))\right)
- g\left((\varphi (-\overline{P2})) \cdot (\overline{\gamma}\cdot\overline{\varepsilon}(P3))\cdot\overline{\gamma}^7\cdot(\varphi (\overline{P1}))\right) \} \\
\{ \{ 1 \to 1, 2 \to 2, 3 \to 3, 4 \to 4 \} \}
\end{pmatrix}
\]
\end{mathematicaNotebook}

\vspace{0.2cm}
\noindent
In this case, there is only one diagram, and the only existing permutation is the identity. 

If we consider now another process, running the input

\begin{mathematicaNotebook}
\begin{mmaCell}[moredefined={modelPhys}]{Input}
  AmplitudeComputation[\{F[1],F[1],-F[1],-F[1]\},6,modelPhys,SeparateCrossings->True] 
\end{mmaCell}
\end{mathematicaNotebook}

\noindent
generates a result similar to the previous one, but now the first element of the output is a list of the amplitudes associated with the two different diagrams we can draw at tree level for this process:

\begin{equation}
 \vcenter{\hbox{\includegraphics[height=2.04cm]{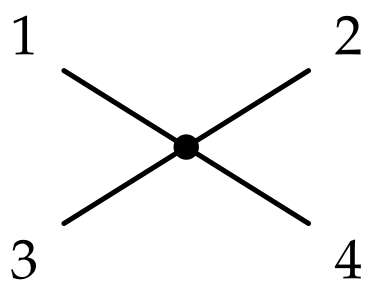}}} +  \left( ~~ \vcenter{\hbox{\includegraphics[height=2cm]{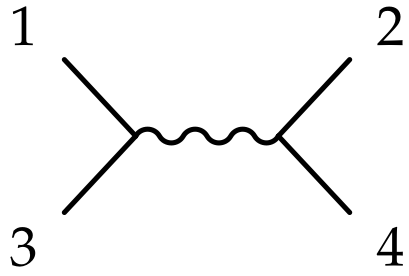}}} +  \vcenter{\hbox{\includegraphics[height=1.98cm]{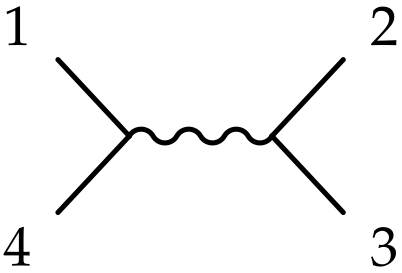}}} ~~ \right)= A_1 + \Big( A_2 + 3 \leftrightarrow 4 \Big)~.
    \label{eq: crossings}
\end{equation}

\vspace{0.3cm}

\noindent
These correspond to the four-fermion contact interaction and the diagram with a gauge boson propagating in an internal line. The second element of the output is a list containing the different allowed permutations of external legs for each diagram. For the contact interaction, there are no permutations. However, for the second diagram, the amplitude 

\[
\left\{- \sum_{i=6,7} \sum_{j=6,7} \frac{g^2 (\varphi (-\overline{P3})) \cdot \overline{\gamma}^{\mathrm{Lor2}} \cdot \overline{\gamma}^i \cdot (\varphi (\overline{P1})) (\varphi (-\overline{P4})) \cdot \overline{\gamma}^{\mathrm{Lor2}} \cdot \overline{\gamma}^j \cdot (\varphi (\overline{P2}))}{(\overline{P2} + \overline{P4})^2} \right\} ~~,
\]

\noindent
calculated using \mmaInlineCell[moredefined=modelPhys]{Input}{modelPhys} ($r_i \to 0$), admits two possible permutations, \textit{i.e.}, the identity and transposing particles 3 and 4:

\[
\{\{ 1 \to 1, 2 \to 2, 3 \to 3, 4 \to 4 \}, \{ 1 \to 1, 2 \to 2, 3 \to 4, 4 \to 3 \}\}~~.
\]

\noindent
The output displayed in \mmaInlineCell{Input}{TraditionalForm} within the 
\mathematica \space notebook appears as follows:

\[
\begin{pmatrix}
\{ A_1 \} && \{A_2 \}\\
\{\{ 1 \to 1, 2 \to 2, 3 \to 3, 4 \to 4 \}\} && 
\{\{ 1 \to 1, 2 \to 2, 3 \to 3, 4 \to 4 \}, \{ 1 \to 1, 2 \to 2, 3 \to 4, 4 \to 3 \}\}
\end{pmatrix} ~~,
\]
where $A_1$ and $A_2$ are the amplitudes corresponding to each diagram.

\noindent If we specify the option \mmaInlineCell{Input}{ExcludeParticles\(\,\,\)->\(\,\,\)\{V[1]\}}, the second diagram is omitted, resulting in the following output:

 \[
\begin{pmatrix}
\{ A_1 \} \\
\{\{ 1 \to 1, 2 \to 2, 3 \to 3, 4 \to 4 \}\} 
\end{pmatrix} ~~.
\]

For convenience, we can store the computed amplitudes and their associated permutations in separate lists using the following command:

\begin{mathematicaNotebook}
\begin{mmaCell}[moredefined={process, model}]{Input}
  \{amps,perms\}=AmplitudeComputation[process,EFTOrder,model,SeparateCrossings->True] 
\end{mmaCell}
\end{mathematicaNotebook}

\noindent
The function
\begin{functiondef}
\mmaInlineCell[moredefined={amps,perms,fermionList}]{Input}{RecoverFullAmplitude[amps,\(\,\,\)perms,\(\,\,\)fermionList]}
\end{functiondef}

\noindent
applies the specified permutations to the partial amplitudes, generating a list of amplitudes with a length equal to the number of existing diagrams. For instance, in the case of equation \eqref{eq: crossings}, this list contains three elements. The function takes, as a third argument, a list of positive integers identifying which external lines are fermions, in order to correctly take into account ``minus'' relative signs between odd and even permutations. If the last argument is an empty list or it is omitted, it will be considered that there are no fermions as external particles. The total amplitude can then be reconstructed by executing the following command:

\begin{mathematicaNotebook}
\begin{mmaCell}[moredefined={amps,perms}]{Input}
  finalAmp = Plus @@ RecoverFullAmplitude[amps, perms]

\end{mmaCell}
\end{mathematicaNotebook}

\subsection{Finding Isomorphisms}

In \texttt{FeynArts}, there are several levels in the representation of a diagram. To find isomorphisms, \tool \space focuses on the most basic level: \textit{topologies}. A topology is simply a set of lines (propagators) connecting a set of points (vertices), where no information about particles and Feynman rules is encoded. This is essentially a graph, so we can make use of graph theory tools to identify such isomorphisms. 

To make the notation clear we briefly expose how we deal with diagrams and graphs. We can uniquely identify a graph with a set of pairs $\{\{v_1,v_2\},\{v_3,v_4\},\dots\}$, where each $v_i$ represents a vertex and each pair a line connecting both vertices \footnote{If the graph is \textit{undirected} the order of the pairs is unimportant.}. In this graph representation of a diagram, external lines are those which are connected to a vertex of degree 1~\footnote{In graph theory, the degree of a vertex is the number of edges that are connected to that vertex.}. Then, every vertex of degree 1 unambiguously identifies an external leg (i.e., an external particle). A permutation of a graph is essentially a permutation over the set $\{v_1,v_2,\dots\}$, as long as it does not exchange vertices of different degrees. This way, a permutation of a graph reliably represents a \textit{crossing} of external particles in a diagram. Finally, two graphs will be called to be \textit{isomorphic} if there exists a permutation that brings one into another, which is nothing but the mathematical way to express that two diagrams are related by a crossing permutation of the external particles.

The function 

\begin{functiondef}
\mmaInlineCell[moredefined={topologyList, allowedPermsList}]{Input}{FindIsomorphisms[topologyList, allowedPermsList]}
\end{functiondef}

\noindent
is designed to identify and classify isomorphic topologies while incorporating constraints on allowable permutations of external legs. It refines the identification of isomorphisms by considering only physically meaningful permutations ---meaning those that only permute identical particles---, ensuring a precise classification of equivalent diagrams via graph representation theory.  

In its initial approach, the function \mmaInlineCell[moredefined=topologyList]{Input}{FindIsomorphisms[topologyList]} classifies isomorphic graphs without considering permutation constraints, that is, taking all external lines to represent the same particle. It does so by comparing the canonical representations of the diagrams, as isomorphic graphs share the same canonical form~\footnote{In graph theory, the canonical form of a graph is a standardized way of representing the graph so that structurally identical graphs (called isomorphic graphs) have the same representation, regardless of how their vertices are labeled. In \mathematica, canonical forms are computed by default using the Bliss algorithm.}. This process identifies a set of representative unique graphs, their corresponding isomorphism classes, and the permutations that map isomorphic graphs to one another. 

Given a canonical diagram $D_0$, its isomorphic graphs or, in group-theory terms, its equivalence class, consists of all the diagrams $\{D_i\}$ satisfying $g_i D_0 = D_i$, where $ g_i$ represents any permutation over external legs. Notice that if the diagrams have $n$ external legs, $g_i$ will be in general an element of $S_n$. 

The next step consist on the further division of equivalence classes attending to the symmetries permitted by \mmaInlineCell[moredefined=allowedPerms]{Input}{allowedPerms}. This argument should be a list of lists of numbers with the form \texttt{\{\{1,\dots,i\}, \{i+1,\dots,j\},\dots,\{k,\dots,n\}\}}, stating that the external legs 1 to $i$ are identical, equally for the legs $i+1$ to $j$ and so on. With this, we aim to restrict the equivalence classes of diagrams related by a $S_n$ permutation to those related by an element of $H = S_{\{1,\dots,i\}} \times S_{\{i+1,\dots,j\}} \times \ldots \times S_{\{k,\dots,n\}}$, where $S_I$ refers to the group of possible permutations over the set $I$. 


However, additional considerations are necessary due to graph automorphisms, which introduce extra symmetries that impact the classification. These are all permutations of a graph which leave the graph invariant, which turn out to form a group $A$. In particular, there exists a degree of freedom in first rotating the representative graph $D_0$ by an arbitrary automorphism $a_0\in A_0$. This means that in the initial classification, we store the permutation $g_i$ for each diagram in the equivalence class, but, unfortunately, it is not unique. Specifically, since  
\[
D_i = g_i D_0 = g_i a^{(i)}_0 D_0 = g'_i D_0,
\]  
it follows that $g_i$ is only determined up to automorphisms $a^{(i)}_0\in A_0$. 

To account for this, we must refine the isomorphism classes further. Given two diagrams $D_i$ and $D_j$ within the equivalence class of $D_0$, they are considered truly equivalent if there exists an element $h \in H$ such that  
\[
h D_i = D_j.
\]  
\vspace{-5pt}
Expanding this condition, we obtain:  
\begin{equation*}
    h D_i = D_j \iff h g_i a^{(i)}_0 D_0 = g_j a^{(j)}_0 D_0 \iff h g_i a^{(i)}_0 = g_j a^{(j)}_0.
\end{equation*}  
Rearranging, and considering that $a_0^{(n)}$ are elements of a group, we get that
\begin{equation}
    g_j = h g_i a^{(i)}_0 \left(a^{(j)}_0\right)^{-1} = h g_i a^{(k)}_0 .
    \label{eq:equivalence relation}
\end{equation}  
From this, we read that we must find equivalence subclasses within $\{g_i\}$ by means of the equivalence relation \eqref{eq:equivalence relation}. This process effectively constructs equivalence classes using \textit{double cosets} $HgA_0$, which account for the symmetries induced by two subgroups: the right cosets of the subgroup of allowed permutations and the left cosets of the automorphism group of the representative diagram.

In practice, we first gather every isomorphic graph no matter the permutation, then take a representative of each class and subdivide the class further attending to the double coset $HgA_0$, where $H$ is the subgroup of allowed permutations and $A_0$ the automorphism group of the representative diagram. Once the refined isomorphism classes are obtained, the function determines new representative graphs for each subclass and computes the permutations required to map graphs onto their respective representatives. Finally, it translates the results back into permutation notation, ensuring that the output remains interpretable in terms of external leg exchanges. The function returns a tuple consisting of the indices of representative graphs, the full classification of graphs, and the corresponding permutations for each isomorphism class.  

To see an explicit example, let us identify the isomorphic classes in the tree-level topologies of $n=8$ external legs with every possible vertex degree from 3 to 8. These are 39$\,$208 different topologies and takes about $5.3$ seconds in this case.

\begin{mathematicaNotebook}
\begin{mmaCell}{Input}
  \{time, topo\} = CreateTopologies[0, 8->0] // Timing;
  time
  topo // Length
    
\end{mmaCell}
\begin{mmaCell}{Output}
  5.28852
  39\(\,\)208
  
\end{mmaCell}
\end{mathematicaNotebook}

\noindent
If we run 

\begin{mathematicaNotebook}
\begin{mmaCell}[moredefined={topo}]{Input}
  \{reprDiag, classDiags, perms\} = FindIsomorphisms[topo, \{\{1,2,3,4,5,6,7,8\}\}];
  
\end{mmaCell}
\end{mathematicaNotebook}

\noindent
we get the list of representative diagrams in \mmaInlineCell[moredefined=topo]{Input}{topo}, the list of classes gathering all equivalent diagrams and the list of permutations among them. In this example we set all external particles to be the same, so every permutation is allowed (in this case, we could have omitted the second argument of \mmaInlineCell{Input}{FindIsomorphisms[]} as well). Graphs in \mmaInlineCell[moredefined=reprDiag]{Input}{reprDiag} and \mmaInlineCell[moredefined=classDiags]{Input}{classDiags} will be identified with integers, corresponding to their position in the \mmaInlineCell{Input}{TopologyList} object \mmaInlineCell[moredefined=topo]{Input}{topo}. Thus, the 39$\,$208 graphs are divided into 32 isomorphic classes whose representatives are the following diagrams:

\begin{mathematicaNotebook}
\begin{mmaCell}[moredefined={reprDiag}]{Input}
  reprDiag // Length
  reprDiag
  
\end{mmaCell}
\begin{mmaCell}{Output}
  32
  \{1, 2, 9, 30, 121, 126, 127, 131, 243, 709, 2\(\,\)039, 2\(\,\)042, 2\(\,\)044, 2\(\,\)046, 2\(\,\)063, 
  2\(\,\)092, 2\(\,\)110, 2\(\,\)111, 2\(\,\)303, 3\(\,\)267, 11\(\,\)489, 11\(\,\)490, 11\(\,\)496, 11\(\,\)513, 11\(\,\)514, 11\(\,\)546, 
  11\(\,\)548, 11\(\,\)828, 28\(\,\)814, 28\(\,\)817, 28\(\,\)819, 28\(\,\)895\}

\end{mmaCell}
\end{mathematicaNotebook}

\noindent
For instance, there are 27 diagrams isomorphic to diagram \#2, occuppying the positions 3 to 8 and 100 to 120 in \mmaInlineCell[moredefined=topo]{Input}{topo}. We can also also check the external leg permutations mapping these diagrams to diagram \#2:

\begin{mathematicaNotebook}
\begin{mmaCell}[moredefined={classDiags,perms}]{Input}
  classDiags\partL2\partR
  perms\partL2\partR // Short
    
\end{mmaCell}
\begin{mmaCell}{Output}
  \{2, 3, 4, 5, 6, 7, 8, 100, 101, 102, 103, 104, 105, 106, 107, 108, 109, 110, 
  111, 112, 113, 114, 115, 116, 117, 118, 119, 120\}
  
\end{mmaCell}
\begin{mmaCell}[form=Short]{Output}
  \{ <|1\(\to\)1,\(\,\,\,\)2\(\to\)2,\(\,\,\,\)3\(\to\)3,\(\,\,\,\)4\(\to\)4,\(\,\,\,\)5\(\to\)5,\(\,\,\,\)6\(\to\)6,\(\,\,\,\)7\(\to\)7,\(\,\,\,\)8\(\to\)8|>, 
  <|1\(\to\)1,\(\,\,\,\)2\(\to\)3,\(\,\,\,\)3\(\to\)2,\(\,\,\,\)4\(\to\)4,\(\,\,\,\)5\(\to\)5,\(\,\,\,\)6\(\to\)6,\(\,\,\,\)7\(\to\)7,\(\,\,\,\)8\(\to\)8|>, <<24>>,
  <|1\(\to\)6,\(\,\,\,\)2\(\to\)8,\(\,\,\,\)3\(\to\)1,\(\,\,\,\)4\(\to\)2,\(\,\,\,\)5\(\to\)3,\(\,\,\,\)6\(\to\)4,\(\,\,\,\)7\(\to\)5,\(\,\,\,\)8\(\to\)7|>,
  <|1\(\to\)7,\(\,\,\,\)2\(\to\)8,\(\,\,\,\)3\(\to\)1,\(\,\,\,\)4\(\to\)2,\(\,\,\,\)5\(\to\)3,\(\,\,\,\)6\(\to\)4,\(\,\,\,\)7\(\to\)5,\(\,\,\,\)8\(\to\)6|> \}
    
\end{mmaCell}
\begin{mmaCell}[moredefined={topo, classDiags},moregraphics={moreig={scale=.5}}]{Input}
  Paint[topo\partL\(\,\)classDiags\partL2\partR\(\,\)\partR, ColumnsXRows -> 7];
  
  \mmaGraphics{Images/topo2}
  
\end{mmaCell}
\end{mathematicaNotebook}

Notice the difference in the result when not all external particles are identical. Indeed, if we intend to identify isomorphisms of diagrams in the process $\psi \overline\psi \eta \eta \phi \phi \phi \phi$, where only crossings between the $\eta$'s (particles 3 and 4) or between the $\phi$'s (particles 5 to 8) are allowed, we now have 

\begin{mathematicaNotebook}
\begin{mmaCell}[moredefined={topo}]{Input}
  \{reprDiag2,classDiags2,perms2\}\(\,\,\)=\(\,\,\)FindIsomorphisms[topo,\(\,\)\{\{1\},\{2\},\{3,4\},\{5,6,7,8\}\}];
  
\end{mmaCell}
\end{mathematicaNotebook}

\noindent
where \mmaInlineCell[moredefined=reprDiag2]{Input}{reprDiag2} contains this time $1\,951$ different diagrams. The 28 diagrams which where isomorphic before (under a general permutation of $S_8$) are now divided in 9 different classes when the equivalence relation is restricted to permutations of $S_{\{3,4\}}\times S_{\{5,6,7,8\}}$. In this particular example, these classes are found from the 2nd to the 10th position of \mmaInlineCell[moredefined=classDiags2]{Input}{classDiags2}, grouping together the following diagrams: $\{2\},\{3,4\},\{5,6,7,8\},\{9,10\},\{100,101\},\{102,103,104,105\},\{106\},\{107,\dots, 114\}$ and $\{115,\dots,120\}$.

Once the topologies are divided, we have a very reduced number of diagram topologies to compute. These bare topologies are afterwards dressed with the proper field content, as usual, with the \mmaInlineCell{Input}{InsertFields[]} function from \texttt{FeynArts}. The advantage of performing the identification of isomorphisms before the dressing is that there is no need to worry about internal lines having the same field content, the information about what external lines can be crossed is enough. If there are more than one way to dress the same topology with different fields, then \mmaInlineCell{Input}{InsertFields[]} will do the job, but the crossing permutations among the different diagrams remain identical. This very process is also automated by the function
\begin{functiondef}
    \mmaInlineCell[moredefined={process,model}]{Input}{DiagramComputation[process, model, ExcludeParticles->\{\}]}
\end{functiondef}
\noindent
which gives a list of two components. The first one is the list of Feynman diagrams which are inequivalent under permutations. Each entry is conformed by a \mmaInlineCell{Input}{TopologyList[]} object and corresponds to one different topology, but it may contain more than one diagram if several distinct particles could be inserted therein. The second component returned is the list of all necessary permutations to each entry of the diagrams list in order to recover the whole diagram set. The function \mmaInlineCell{Input}{DiagramComputation[]} takes care of computing the necessary topologies and applying \mmaInlineCell{Input}{FindIsomorphisms[]}, followed by the \mmaInlineCell{Input}{InsertFields[]} command.

As a final comment before leaving this subsection, notice that \mmaInlineCell{Input}{FindIsomorphisms[]} is equally suitable for loop diagrams, correctly identifying crossings at the loop level. However, given the lack of knowledge about fermion flow at the topology level, with closed fermion loops there may be some diagrams which turn out to be related by a crossing permutation but they cannot be identified as such. This is because in \texttt{FeynArts} they correspond to the exact same topology, later becoming different diagrams with opposite fermion flow after the \mmaInlineCell{Input}{InsertFields[]} command. For instance, this would happen with the closed fermion triangle loop diagrams, in which transposing two of the external legs is equivalent to the change of the direction of the fermion flow.

\subsection{Rational kinematics}

Once amplitudes are computed we have to replace every kinematic element with a numerical value. In the end, we expect every amplitude to be a complex polynomial in the WCs with, possibly, rational functions of the masses of the fields. Furthermore, we want to ensure that coefficients are rational so that no machine precision is lost over the whole process and the matching is exact. 

To this end, we can exploit the \textit{momentum twistor} theory introduced in \cite{Hodges:2009hk} to generate rational-valued momenta satisfying momentum conservation and on-shell conditions, as shown in \cite{DeAngelis:2022qco}. This is implemented in the \texttt{Mathematica} package found in \url{https://github.com/StefanoDeAngelis/SpinorHelicity/tree/main/Wolfram/NumericalKinematics.wl}, which we provide in the \texttt{mosca/ ExternalPackages/NumericalKinematics/} folder with a slight modification to allow for the generation of momenta with symbolic masses $m_i$. 

On the other hand, in the \texttt{KinematicSubstitution.m} file we implemented routines to transform the output of \texttt{NumericalKinematics.wl} to facilitate the substitution of such rational kinematics into amplitudes. One of the main functions is 
\begin{functiondef}
\mmaInlineCell[moredefined={nF, nV, nS, masses}]{Input}{KinematicConfigurations[nF, nV, nS, masses]}
\end{functiondef}
\noindent
which generates a configuration of momenta, Dirac spinors and polarizations for $nF$ fermions, $nV$ vectors and $nS$ scalars with masses $\{m_1,\dots,m_{nF+nV+nS}\}$, where the first $nF$ masses correspond to fermions and so on. Here, we recall that we do not address kinematic configurations for massive vectors, so the corresponding vector masses entries should be filled with 0's. The argument \texttt{masses} is to be provided with the list of masses, but alternatively also accepts either a single symbol (or number), in which case every particle is considered to have the same mass, or no argument whatsoever and then every particle is considered to be massless. 

The output of \mmaInlineCell{Input}{KinematicConfigurations[]} is an association. We can access the list of momenta, $u$ Dirac spinors, $\bar v$ Dirac spinors, $\epsilon^+$ positive polarizations and $\epsilon^-$ negative polarizations through the entries \texttt{"p", "u", "vbar", "e+"} and \texttt{"e-"}, respectively \footnote{Since we consider both fermions and antifermions always incoming, we refer to spinors as $u$ and to conjugate spinors as $\bar v$. However, what we call here $\bar v$ would actually be a $\bar u$ in case of an outgoing fermion and \textit{vice versa}.}. Each of these is, in turn, another association where we have a unique key \texttt{fi,vi,si} for the $i$-th fermion, vector or scalar, respectively. Thereby, for instance, the momentum of the $i$-th scalar will be obtained through the entries \texttt{\{"p",si\}}, and the spinor of the $j$-th fermion through \texttt{\{"u",fj\}}.

To see an explicit example, let us generate a kinematic configuration with two fermions with masses $m_1,m_2$ and a massless vector.

\begin{mathematicaNotebook}
\begin{mmaCell}[moredefined={m1,m2}]{Input}
  kin = KinematicConfigurations[2, 1, 0, \{m1, m2, 0\}];
  
\end{mmaCell}
\end{mathematicaNotebook}

\noindent
The momentum (a four-vector) for the first fermion takes the following form:

\begin{mathematicaNotebook}
\begin{mmaCell}[moredefined={kin}]{Input}
  kin["p",f1]
  
\end{mmaCell}
\begin{mmaCell}{Output}
  \{ -\mmaFrac{43831}{47824}\,\,-\,\,\mmaFrac{4\,\,\mmaSup{m1}{2}}{61} , \mmaFrac{77459}{95648}\,\,-\,\,\mmaFrac{40\,\,\mmaSup{m1}{2}}{61} , -\mmaFrac{I(8429\,\,+\,\,62720\,\,\mmaSup{m1}{2})}{
  95648} , -\mmaFrac{2993}{6832}\,\,-\,\,\mmaFrac{4 \mmaSup{m1}{2}}{61} \}
  
\end{mmaCell}
\end{mathematicaNotebook}

\noindent
To take the list of all momenta one can use the command \mmaInlineCell{Input}{Values}. We can check momentum conservation and that every momentum is on-shell by running

\begin{mathematicaNotebook}
\begin{mmaCell}[moredefined={kin}]{Input}
  plist = Values @ kin["p"];
  Plus @@ plist // Simplify
  (MDot[#, #] \& /@ plist) // Simplify
  
\end{mmaCell}
\begin{mmaCell}{Output}
  \{ 0 , 0 , 0 , 0 \}
  \{ \mmaSup{m1}{2} , \mmaSup{m2}{2} , 0 \}
  
\end{mmaCell}
\end{mathematicaNotebook}

\noindent
Here, \mmaInlineCell[moredefined={a,b}]{Input}{MDot[a,b]} is the Minkowsi product $a\cdot b$ of the two four-vectors $a$ and $b$.

Furthermore, Dirac equations for spinors, $(\slashed{p} - m) u = 0$, and for conjugate spinors, $\bar v(\slashed{p} + m) = 0$, are satisfied:

\begin{mathematicaNotebook}
\begin{mmaCell}[moredefined={kin}]{Input}
  (pslash[kin["p",f1]]\,\,.\,\,kin["u",f1])\,\,-\,\,m1\,\,kin["u",f1] // Simplify
  (kin["vbar",f1]\,\,.\,\,pslash[kin["p",f1]]) + m1 kin["vbar",f1] // Simplify
  
\end{mmaCell}
\begin{mmaCell}{Output}
  \{ 0 , 0 , 0 , 0 \}
  \{ 0 , 0 , 0 , 0 \}
  
\end{mmaCell}
\end{mathematicaNotebook}

\noindent
where \mmaInlineCell[moredefined=a]{Input}{pslash[a]} ``slashes'' the four-vector $a$ by applying $\slashed{a}=\sum_{\mu=0}^3 \gamma^\mu a_\mu$. Finally, both polarizations are also transverse

\begin{mathematicaNotebook}

\begin{mmaCell}[moredefined={kin}]{Input}
  MDot[kin["e+",v1], kin["p",v1]] // Simplify
  MDot[kin["e-",v1], kin["p",v1]] // Simplify
  
\end{mmaCell}
\begin{mmaCell}{Output}
  0
  0
  
\end{mmaCell}

\end{mathematicaNotebook}
\noindent
and consist of light-like vectors verifying $\upepsilon^+\cdot\upepsilon^-=-1$:

\begin{mathematicaNotebook}

\begin{mmaCell}[moredefined={kin}]{Input}
  MDot[kin["e+",v1], kin["e+",v1]] // Simplify
  MDot[kin["e-",v1], kin["e-",v1]] // Simplify
  MDot[kin["e+",v1], kin["e-",v1]] // Simplify
  
\end{mmaCell}
\begin{mmaCell}{Output}
  0
  0
  -1
  
\end{mmaCell}

\end{mathematicaNotebook}

The second important command is 

\begin{functiondef}
\mmaInlineCell[moredefined={amp, nF, nV, nS, masses}]{Input}{ReplaceKinematics[amp, nF, nV, nS, masses, Momenta\(\,\,\)->\(\,\,\)\{\},}
\mmaInlineCell{Input}{PolarizationInfo\(\,\,\)->\(\,\,\)True/False]}
\end{functiondef}

\noindent
This function is designed to receive an amplitude \mmaInlineCell[moredefined=amp]{Input}{amp} (or a list of amplitudes) and replace specific random rational kinematics in it. The format for specifying the number of particles and their respective masses is identical to the one in \mmaInlineCell{Input}{KinematicConfigurations[]}. \mmaInlineCell{Input}{ReplaceKinematics[]} effectively replaces values for momenta and polarization products, contractions with the Levi-Civita tensor $\varepsilon_{\mu\nu\rho\sigma}$ and Dirac chains, thus transforming the amplitude into a complex polynomial over the WCs (admitting rational functions of the masses though). The \mmaInlineCell{Input}{Momenta} option accepts a list which allows to specify, in order, the symbol for the momenta of the fermions, vectors and scalars in \mmaInlineCell[moredefined=amp]{Input}{amp}. If no option is given, it will be assumed that the $n$-th particle has momentum \mmaInlineCell[moredefined=Pn]{Input}{Pn}. 

Let us see how it works by computing a numerical value for the following expression:
$$
g \,\, \overline v_1 \frac{\slashed{q}_1-\slashed{q}_2}{(q_1-q_2)^2} u_2 \,,
$$
where we have an antifermion and a fermion with momenta $p_1,p_2$, respectively, and 2 scalars with momentum $q_1,q_2$. Moreover, the fermions are massless and the scalar with momentum $q_i$ has mass $m_i$. Then, one would have to type the following: 

\begin{mathematicaNotebook}
\begin{mmaCell}[moredefined={SpinorVBar,GS,SP,SpinorU}]{Input}
  amp = g SpinorVBar[p1]\(\,\,\,\).\(\,\,\,\)(GS[q1-q2]\(\,\,\,\)/\(\,\,\,\)SP[q1-q2, q1-q2])\(\,\,\,\).\(\,\,\,\)SpinorU[p2];
  ReplaceKinematics[amp, 2, 0, 2, \{0,0,m1,m2\}, Momenta->\{p1,p2,q1,q2\}] // Simplify
    
\end{mmaCell}
\begin{mmaCell}{Output}
  \{ -\(\,\,\,\)\mmaFrac{g\(\,\,\,\)(447219682\(\,\,\,\)+\(\,\,\,\)172476999\(\,\,\,\)\mmaSup{m1}{2}\(\,\,\,\)+\(\,\,\,\)194337\(\,\,\,\)\mmaSup{m2}{2})}{195\(\,\,\,\)(42343\(\,\,\,\)+\(\,\,\,\)14205\(\,\,\,\)\mmaSup{m1}{2}\(\,\,\,\)+\(\,\,\,\)1661 \mmaSup{m2}{2})} \}
    
\end{mmaCell}
\end{mathematicaNotebook}

If vectors are present, then several numerical amplitudes will be returned, attending to the different combinations of $+$ and $-$ helicity polarizations. To see which amplitude comes from which polarization choice one can simply set the \mmaInlineCell{Input}{PolarizationInfo} option to \mmaInlineCell{Input}{True} (which is in \mmaInlineCell{Input}{False} by default). For instance, we can compute the 4 gluon amplitude with \texttt{FeynArts+FeynCalc} and replace a physical configuration of momenta:

\begin{mathematicaNotebook}
\begin{mmaCell}{Input}
  ts = CreateTopologies[0, 4 -> 0];
  diags\(\,\,\,\)=\(\,\,\,\)InsertFields[ts,\(\,\,\,\)\{V[1],V[1],V[1],V[1]\}\(\,\,\)->\(\,\,\)\{\},\(\,\,\,\)InsertionLevel\(\,\,\)->\(\,\,\)\{Particles\},
          Model\(\,\,\)->\(\,\,\)modelGluon, GenericModel\(\,\,\)->\(\,\,\)modelGluon];
  amp = FCFAConvert[CreateFeynAmp[diags], IncomingMomenta\(\,\,\)->\(\,\,\)\{P1,P2,P3,P4\}, 
            List\(\,\,\)->\(\,\,\)False, DropSumOver\(\,\,\)->\(\,\,\)True];
  amp = FeynAmpDenominatorExplicit[SUNSimplify[amp, SUNFJacobi\(\,\,\)->\(\,\,\)True]];
  ReplaceKinematics[amp,\(\,\,\,\)0,\(\,\,\,\)4,\(\,\,\,\)0,\(\,\,\,\)0,\(\,\,\,\)PolarizationInfo\(\,\,\)->\(\,\,\)True] // Simplify
    
\end{mmaCell}
\begin{mmaCell}{Print}
  Polarization(s) for P1, P2, P3, P4 read: \{++++, +++-, ++-+, ++--, +-++, +-+-, 
  +--+, +---, -+++, -++-, -+-+, -+--, --++, --+-, ---+, ----\}
  
\end{mmaCell}
\begin{mmaCell}{Output}
  \{0,\(\,\,\,\)0,\(\,\,\,\)0,\(\,\,\,\)\mmaFrac{768}{551}\mmaSubSup{g}{s}{2}(57\(\,\,\)\mmaSup{f}{g1g4a}\mmaSup{f}{g2g3a}\(\,\,\)+\(\,\,\)\mmaSup{f}{g1g3a}\mmaSup{f}{g2g4a}),\(\,\,\,\)0,\(\,\,\,\)\mmaFrac{3888}{551}\mmaSubSup{g}{s}{2}(57\(\,\,\)\mmaSup{f}{g1g4a}\mmaSup{f}{g2g3a}\(\,\,\)+\(\,\,\)\mmaSup{f}{g1g3a}\mmaSup{f}{g2g4a}),
  \mmaFrac{3}{8816}\mmaSubSup{g}{s}{2}(57\(\,\,\)\mmaSup{f}{g1g4a}\mmaSup{f}{g2g3a}\(\,\,\)+\(\,\,\)\mmaSup{f}{g1g3a}\mmaSup{f}{g2g4a}),\(\,\,\,\)0,\(\,\,\,\)0,\(\,\,\,\)\mmaFrac{16}{14877}\mmaSubSup{g}{s}{2}(57\(\,\,\)\mmaSup{f}{g1g4a}\mmaSup{f}{g2g3a}\(\,\,\)+\(\,\,\)\mmaSup{f}{g1g3a}\mmaSup{f}{g2g4a}),
  \mmaFrac{6859}{12528}\mmaSubSup{g}{s}{2}(57\(\,\,\)\mmaSup{f}{g1g4a}\mmaSup{f}{g2g3a}\(\,\,\)+\(\,\,\)\mmaSup{f}{g1g3a}\mmaSup{f}{g2g4a}),\(\,\,\,\)0,\(\,\,\,\)\mmaFrac{24389}{8208}\mmaSubSup{g}{s}{2}(57\(\,\,\)\mmaSup{f}{g1g4a}\mmaSup{f}{g2g3a}\(\,\,\)+\(\,\,\)\mmaSup{f}{g1g3a}\mmaSup{f}{g2g4a}),\(\,\,\,\)0,\(\,\,\,\)0,\(\,\,\,\)0\}
  
\end{mmaCell}
\end{mathematicaNotebook}

\noindent
We can see that the only non-zero amplitudes are those having two gluons of the same helicity and the other two with the other one, in agreement with the very well-known fact that in an $n$-gluon amplitude at least two gluons need to have different helicity from the rest to give a non-vanishing result. Since every particle is massless, here we called \mmaInlineCell{Input}{ReplaceKinematics[]} with just a 0 in the fifth argument, instead of a list of four 0's.

We finally comment some aspects. First, notice that, in the same call to \mmaInlineCell{Input}{ReplaceKinematics[]}, the momenta is not being randomly re-computed in each case; it is computed once and used for every helicity configuration instead. This makes this function very useful for computing numerical values for cross-sections with sums over helicities. Second, for fermions we do not have distinction between $+1/2$ and $-1/2$ helicities because we consider all fermions to be Dirac. However, we can equally mimic configurations for Weyl fermions with well-defined helicites by adding the corresponding $P_{L/R}$ projectors in the spinor chains. Finally, if we want to replace the same rational kinematic configuration in several amplitudes, we can just provide them in a list as the first argument of \mmaInlineCell{Input}{ReplaceKinematics[]}.

\subsection{Green's Basis Reduction}

The main function of \tool \space is

\begin{functiondef}
\mmaInlineCell[moredefined={model1,model2,OutputFormat}]{Input}{RedBasis[model1, model2, EFTOrder, OutputFormat\(\,\,\)->\(\,\,\)\{\}, ExcludeParticles\(\,\,\)->\(\,\,\)\{\},}
\mmaInlineCell{Input}{StopAt\(\,\,\)->\(\,\,\)\{\}]}
\end{functiondef}

\noindent
This function returns the redefinition of the WCs of \mmaInlineCell[moredefined=model2]{Input}{model2} in terms of the WCs of \mmaInlineCell[moredefined=model1]{Input}{model1} up to the dimension fixed by \mmaInlineCell{Input}{EFTOrder}. The function also accepts three optional parameters: (1) The \mmaInlineCell{Input}{OutputFormat} option provides control over the display format of the result by specifying new values for \mmaInlineCell{Input}{ModelName}, \mmaInlineCell{Input}{VisibleModel} and \mmaInlineCell{Input}{VisibleOrder}. If \mmaInlineCell{Input}{OutputFormat} is left as an empty list, the output will follow the formatting settings previously defined through the \mmaInlineCell{Input}{FormatWC[]} function. (2) The \mmaInlineCell{Input}{ExcludeParticles} option, which defaults to an empty list, allows the user to exclude any particle present in the models from the calculation. (3) The \mmaInlineCell{Input}{StopAt} option sets the process at which the computation should stop, if needed. Therefore, understanding how \mmaInlineCell{Input}{RedBasis[]} determines the order of process computation is essential for effectively using this option. In this subsection, in order to give practical examples and see all these commands in action, we will take \mmaInlineCell[moredefined=model1]{Input}{model1} $\to$ \mmaInlineCell[moredefined=modelFull]{Input}{modelFull}, corresponding to the Lagrangian \eqref{eq: Lagrangian full} and \eqref{eq: Lagrangian interaction},  and \mmaInlineCell[moredefined=model2]{Input}{model2}  $\to$  \mmaInlineCell[moredefined=modelPhys]{Input}{modelPhys}, corresponding to the same Lagrangian where all redundant operators (those with WCs $r_i$) are set to zero.

In \mmaInlineCell{Input}{RedBasis[]}, the process list is first reduced by removing redundant processes that share similar Feynman rules for the contact interaction, as they do not provide additional information. Once this minimum set is obtained, the remaining processes are ordered according to specific criteria: priority is given to those with fewer total particles, a higher number of coefficients relative to kinematical structures, fewer distinct kinematical structures, and a greater number of identical particles. From this list, 2-point interactions are also excluded, as their contribution to the reduction is already accounted for by the \mmaInlineCell{Input}{AllPropagatorAttributes[]} function.

The list of processes can be accessed by using the function 

\begin{functiondef}
\mmaInlineCell[moredefined=model]{Input}{ProcessList[model,\(\,\,\)ExcludeParticles\(\,\,\)->\(\,\,\)\{\},\(\,\,\)Sorted\(\,\,\)->\(\,\,\)True/False]}
\end{functiondef}

\noindent
This function returns a list of computable processes within a given model, excluding the particles specified by \mmaInlineCell{Input}{ExcludeParticles}. Since all particles are considered to be incoming, the output consists of lists representing the participating fields in each process. For example, running the command:

\begin{mathematicaNotebook}
\begin{mmaCell}[moredefined={modelPhys}]{Input}
  ProcessList[modelPhys]
\end{mmaCell}
\end{mathematicaNotebook}

\noindent
produces the output:

\begin{mathematicaNotebook}
\begin{mmaCell}{Output}
  \{\{-F[1],\(\,\,\)F[1]\}, \{S[1],\(\,\,\)S[1]\}, \{V[1],\(\,\,\)V[1]\}, \{-F[1],\(\,\,\)F[1],\(\,\,\)V[1]\}, \{-F[1],\(\,\,\)-F[1],
  \(\,\,\)F[1],\(\,\,\)F[1]\}, \{S[1],\(\,\,\)S[1],\(\,\,\)S[1],\(\,\,\)S[1]\}, \{S[1],\(\,\,\)S[1],\(\,\,\)S[1],\(\,\,\)S[1],\(\,\,\)S[1],\(\,\,\)S[1]\}\}
  
\end{mmaCell}
\end{mathematicaNotebook}

\noindent
The  \mmaInlineCell{Input}{Sorted} option allows to output the list of processes used by \mmaInlineCell{Input}{RedBasis[]} during the computation, following the same ordering criteria and excluding 2-point interactions. Having access to this ordered list can be helpful for selecting a specific process at which to stop the calculation in \mmaInlineCell{Input}{RedBasis[]}, using the \mmaInlineCell{Input}{StopAt} option. For the previous example, setting \mmaInlineCell{Input}{Sorted\(\,\,\)->\(\,\,\)True} results in: 

\begin{mathematicaNotebook}
\begin{mmaCell}[moredefined={modelPhys}]{Input}
  ProcessList[modelPhys, Sorted->True]
\end{mmaCell}
\begin{mmaCell}{Output}
  \{\{-F[1],\(\,\,\)F[1],\(\,\,\)V[1]\}, \{S[1],\(\,\,\)S[1],\(\,\,\)S[1],\(\,\,\)S[1]\}, \{-F[1],\(\,\,\)-F[1],\(\,\,\)F[1],\(\,\,\)F[1]\}, 
  \{S[1],\(\,\,\)S[1],\(\,\,\)S[1],\(\,\,\)S[1],\(\,\,\)S[1],\(\,\,\)S[1]\}\}
  
\end{mmaCell}
\end{mathematicaNotebook}

\mmaInlineCell{Input}{RedBasis[]} makes use of two other functions designed to perform the matching between the amplitudes of two different models: \mmaInlineCell{Input}{MassReduction[]} and \mmaInlineCell{Input}{AmplitudeMatching[]}. 

First,
\begin{functiondef}
    \mmaInlineCell[moredefined={field,model1,model2}]{Input}{MassReduction[field, model1, model2, EFTOrder, Output->True/False]}
\end{functiondef}
\noindent
returns the expression that the bare mass of the given \mmaInlineCell[moredefined=field]{Input}{field} in \mmaInlineCell[moredefined=model2]{Input}{model2} must take, in terms of the bare mass and other 2-point WCs in \mmaInlineCell[moredefined=model1]{Input}{model1}, to require that the physical mass for such field is the same in both models, up to a given EFT order. The \mmaInlineCell{Input}{Output->True} (default) enables to display the result in a readable manner. For instance, by running

\begin{mathematicaNotebook}
\begin{mmaCell}[moredefined={modelFull,modelPhys}]{Input}
  MassReduction[S[1], modelFull, modelPhys, 6]
\end{mmaCell}
\begin{mmaCell}{Print}
  \textbf{Mass redefinition for S[1]:}
  \mmaSubSup{m}{phys}{(4)} \(\to\) \mmaSubSup{m}{full}{(4)}\(\,\,\)-\(\,\,\)\mmaSup{(\mmaSubSup{m}{full}{(4)})}{3}\(\,\,\)\mmaSubSup{rD\(\phi\)}{full}{(6)}
  
\end{mmaCell}
\begin{mmaCell}{Output}
  \{ \mmaSubSup{m}{phys}{(4)} \(\to\) \mmaSubSup{m}{full}{(4)}, \mmaSubSup{m}{phys}{(5)} \(\to\) 0, \mmaSubSup{m}{phys}{(6)} \(\to\) -\mmaSup{(\mmaSubSup{m}{full}{(4)})}{3}\(\,\,\)\mmaSubSup{rD\(\phi\)}{full}{(6)} \}
\end{mmaCell}
\end{mathematicaNotebook}

\noindent
we obtain the physical mass of \mmaInlineCell{Input}{S[1]} ($\phi$) in terms of its bare mass in \mmaInlineCell[moredefined=modelFull]{Input}{modelFull}, up to dimension 6 in the EFT order. For the sake of clarity, we have set both \mmaInlineCell{Input}{VisibleModel\(\,\,\)->\(\,\,\)True} and \mmaInlineCell{Input}{VisibleModel\(\,\,\)->\(\,\,\)True} in the \mmaInlineCell{Input}{FormatWC[]} function to explicitly show the model names and orders in the redefinition. The display option shows the full redefinition of the mass, while the output of the command itself is split order by order in the EFT expansion: $m = m^{(4)} + \frac{1}{\Lambda} m^{(5)} + \frac{1}{\Lambda^2} m^{(6)} + \dots~$. This will be the \textit{modus operandi} in the following commands. One can easily go back and forth between these two representations by applying \mmaInlineCell[moredefined=JoinOrderWC]{Input}{JoinOrderWC[]} and \mmaInlineCell[moredefined=SplitOrderWC]{Input}{SplitOrderWC[]} directly over the replacement rule list.

It is important to notice that here \mmaInlineCell{Print}{\mmaSub{m}{phys}} is \textit{not} the physical mass of $\phi$, but rather the bare mass of $\phi$ appearing in the Lagrangian of \mmaInlineCell[moredefined=modelPhys]{Input}{modelPhys}, whose WCs have been tagged with \mmaInlineCell{Input}{ModelName->"phys"} in the \mmaInlineCell{Input}{FormatWC[]} command. However, in a Lagrangian with no 2-point operators, like the one in \mmaInlineCell[moredefined=modelPhys]{Input}{modelPhys}, the bare mass naturally coincide with the physical mass at tree level, and hence the reduction looks identical to the one obtained when applying \mmaInlineCell{Input}{PropagatorAttributes[]} of $\phi$ in \mmaInlineCell[moredefined=modelFull]{Input}{modelFull} (except for the fact that there we have the squared mass relation).

The function  
\begin{functiondef}
    \mmaInlineCell[moredefined={model1,model2,AllMassReduction}]{Input}
  {AllMassReduction[model1, model2, EFTOrder, Output->True/False,}
  \mmaInlineCell{Input}{ExcludeParticles\(\,\,\)->\(\,\,\)\{\}]}
\end{functiondef}
\noindent
computes the mass redefinition for all fields in \mmaInlineCell[moredefined=model2]{Input}{model2}. The \mmaInlineCell{Input}{Output} option behaves the same as in \mmaInlineCell{Input}{MassReduction[]}, and \mmaInlineCell{Input}{ExcludeParticles} allows to skip any field in the computation. In this specific case we would further have:
\begin{mathematicaNotebook}
\begin{mmaCell}{Print}
  \textbf{Mass redefinition for F[1]:}
  \mmaSub{m\(\psi\)}{phys} \(\to\) \mmaSub{m\(\psi\)}{full}\(\,\,\)+\(\,\,\)\mmaSubSup{m\(\psi\)}{full}{3}\(\,\,\)\mmaSub{r\(\psi\)D}{full}

  \textbf{Mass redefinition for V[1]:}
  0 \(\to\) 0
  
\end{mmaCell}
\end{mathematicaNotebook}

After obtaining the relations of the bare masses of \mmaInlineCell[moredefined=model2]{Input}{model2} in terms of the WCs of \mmaInlineCell[moredefined=model1]{Input}{model1}, it is necessary to find the relations regarding other WCs. The command responsible for this is: 

\begin{functiondef}
\mmaInlineCell[moredefined={model1,model2,process}]{Input}{AmplitudeMatching[model1, model2, process, EFTOrder, ExcludeParticles\(\,\,\)->\(\,\,\)\{\}, }
\mmaInlineCell{Input}{RecalcPropAttributes->\(\,\,\)True/False, Replacements\(\,\,\)->\(\,\,\)\{\{\},\{\}\}]}
\end{functiondef}

\noindent
This function returns the redefinition order by order for all the coefficients in \mmaInlineCell[moredefined=model2]{Input}{model2} contributing to the given process up to the mass dimension fixed by \mmaInlineCell{Input}{EFTOrder}, excluding the particles specified by \mmaInlineCell{Input}{ExcludeParticles}. In some sense, \mmaInlineCell{Input}{AmplitudeMatching[]} is a generalization of \mmaInlineCell{Input}{MassReduction[]} for for processes other than  \mmaInlineCell[moredefined=field]{Input}{field\(\,\,\)->field}. For a matter of convenience, the redefinition will be given in terms of \mmaInlineCell{Input}{MPhysSymbol[]} objects, so the \mmaInlineCell{Input}{ReplacePhysMass[]} function can be useful to obtain a result depending on bare masses of \mmaInlineCell[moredefined=model1]{Input}{model1}. 

\mmaInlineCell[moredefined={RecalcPropAttributes}]{Input}{RecalcPropAttributes} is an option that controls whether 2-point function calculations are repeated. By default, it is set to \mmaInlineCell{Input}{True}. Setting it to \mmaInlineCell{Input}{False} skips this recalculation if the values have already been computed. Caution is advised when disabling this feature, especially when matching calculations are performed at different orders in the EFT expansion. Whenever the EFT order is increased, the propagator attributes must be extended accordingly to ensure consistency of the results. 

On the other hand, the \mmaInlineCell{Input}{Replacements} option, which defaults to a list of two empty lists, allows to apply any replacement rule in the amplitudes coming from \mmaInlineCell[moredefined=model1]{Input}{model1} and \mmaInlineCell[moredefined=model2]{Input}{model2}. This is particularly useful to send some WCs at zero and to substitute known redefinitions for any of the coefficients.

Let us see this at work by considering two processes involving the scalar $\phi$, \mmaInlineCell{Input}{\{S[1],\(\,\,\)S[1],\(\,\,\)S[1]\}} and \mmaInlineCell{Input}{\{S[1],\(\,\,\)S[1],\(\,\,\)S[1],\(\,\,\)S[1],\(\,\,\)S[1],\(\,\,\)S[1]\}}. 
Running \mmaInlineCell{Input}{AmplitudeMatching[]} results in:

\begin{mathematicaNotebook}
\begin{mmaCell}[moredefined={modelFull,modelPhys}]{Input}
  sol1\(\,\,\)=\(\,\,\)AmplitudeMatching[modelFull,\(\,\,\)modelPhys,\(\,\,\)\{S[1],\(\,\,\)S[1],\(\,\,\)S[1],\(\,\,\)S[1]\}, 6]
\end{mmaCell}
\begin{mmaCell}{Output}
  \{\mmaSubSup{lmbd}{phys}{(4)}\(\,\,\)\(\to\)\(\,\,\)\mmaSubSup{lmbd}{full}{(4)},\(\,\,\)\mmaSubSup{lmbd}{phys}{(5)}\(\,\,\)\(\to\)\(\,\,\)0, 
  
   \mmaSubSup{lmbd}{phys}{(6)}\(\,\,\)\(\to\)\(\,\,\)\mmaSup{MPhysSymbol[S[1]]}{2}(-8\(\,\,\)\mmaSubSup{lmbd}{full}{(4)} \mmaSubSup{rD\(\phi\)}{full}{(6)}+ \mmaSubSup{r\(\phi\)\,D}{full}{(6)})\}
  
\end{mmaCell}
\begin{mmaCell}[moredefined={modelFull,modelPhys}]{Input}
  sol2\(\,\,\)=\(\,\,\)AmplitudeMatching[modelFull,\(\,\,\)modelPhys,\(\,\,\)\{S[1],S[1],S[1],S[1],S[1],S[1]\},6]
\end{mmaCell}
\begin{mmaCell}{Output}
  \{\mmaSubSup{lmbd}{phys}{(4)}\(\,\,\)\(\to\)\(\,\,\)\mmaSubSup{lmbd}{full}{(4)},\(\,\,\)\mmaSubSup{lmbd}{phys}{(5)}\(\,\,\)\(\to\)\(\,\,\)0, 

  \mmaSubSup{a\(\phi\)}{phys}{(6)}\(\,\,\)\(\to\)\(\,\,\)\mmaSubSup{a\(\phi\)}{full}{(6)}+ 4\(\,\,\)\mmaSubSup{lmbd}{full}{(4)}(4\(\,\,\)\mmaSubSup{lmbd}{full}{(4)}\mmaSubSup{rD\(\phi\)}{full}{(6)}-\mmaSubSup{r\(\phi\)\,D}{full}{(6)}),
  
  \mmaSubSup{lmbd}{phys}{(6)}\(\,\,\)\(\to\)\(\,\,\)\mmaSup{MPhysSymbol[S[1]]}{2}(-8\(\,\,\)\mmaSubSup{lmbd}{full}{(4)} \mmaSubSup{rD\(\phi\)}{full}{(6)}+ \mmaSubSup{r\(\phi\)\,D}{full}{(6)})\}

\end{mmaCell}
\end{mathematicaNotebook}

\noindent
By applying \mmaInlineCell{Input}{ReplacePhysMass[]} to the result with the following command we obtain:

\begin{mathematicaNotebook}
\begin{mmaCell}[moredefined={modelFull,sol1}]{Input}
  sol1BareMass = (#\partL1\partR\(\,\,\)->\(\,\,\)ReplacePhysMass[#\partL2\partR, 6, modelFull]) \& /@ sol1
\end{mmaCell}
\begin{mmaCell}{Output}
  \{\mmaSubSup{lmbd}{phys}{(4)}\(\,\,\)\(\to\)\(\,\,\)\mmaSubSup{lmbd}{full}{(4)},  \mmaSubSup{lmbd}{phys}{(5)}\(\,\,\)\(\to\)\(\,\,\)0, 
  
   \mmaSubSup{lmbd}{phys}{(6)}\(\,\,\)\(\to\)\(\,\,\)-8\(\,\,\)\mmaSubSup{lmbd}{full}{(4)}\mmaSup{(\mmaSubSup{m}{full}{(4)})}{2}\mmaSubSup{rD\(\phi\)}{full}{(6)}+\mmaSup{(\mmaSubSup{m}{full}{(4)})}{2}\mmaSubSup{r\(\phi\)\,D}{full}{(6)}\}
  
\end{mmaCell}
\begin{mmaCell}[moredefined={sol2,modelFull}]{Input}
  sol2BareMass = (#\partL1\partR\(\,\,\)->\(\,\,\)ReplacePhysMass[#\partL2\partR, 6, modelFull]) \& /@ sol2
\end{mmaCell}
\begin{mmaCell}{Output}
  \{\mmaSubSup{lmbd}{phys}{(4)}\(\,\,\)\(\to\)\(\,\,\)\mmaSubSup{lmbd}{full}{(4)},\(\,\,\)\mmaSubSup{lmbd}{phys}{(5)}\(\,\,\)\(\to\)\(\,\,\)0, 

  \mmaSubSup{a\(\phi\)}{phys}{(6)}\(\,\,\)\(\to\)\(\,\,\)\mmaSubSup{a\(\phi\)}{full}{(6)}+ 16\(\,\,\)\mmaSup{(\mmaSubSup{lmbd}{full}{(4)})}{2}\mmaSubSup{rD\(\phi\)}{full}{(6)}-4\(\,\,\)\mmaSubSup{lmbd}{full}{(4)}\mmaSubSup{r\(\phi\)\,D}{full}{(6)}),
  
  \mmaSubSup{lmbd}{phys}{(6)}\(\,\,\)\(\to\)\(\,\,\)-8\(\,\,\)\mmaSubSup{lmbd}{full}{(4)}\mmaSup{(\mmaSubSup{m}{full}{(4)})}{2}\mmaSubSup{rD\(\phi\)}{full}{(6)}+\mmaSup{(\mmaSubSup{m}{full}{(4)})}{2}\mmaSubSup{r\(\phi\)\,D}{full}{(6)}\}

\end{mmaCell}
\end{mathematicaNotebook}

\noindent
We can see how the first process only matched the $\lambda$ WC but the second one was enough to match both, $\lambda$ and $a_\phi$. We would like to highlight that this, as a general property of on-shell matching, constitutes one key difference with its off-shell counterpart, where one needs to deal with all the amplitudes, one by one, in order to match all WCs. Crucially, like two sides of the same coin, either matching on-shell allows to match many WCs with a unique amplitude, or otherwise it requires a high degree of redundancy which translates into strong cross-checks in the process.

In the latter case, in which one wants to match WCs with amplitudes of increasing number of fields, it is key to use the \mmaInlineCell{Input}{Replacements} option in order to substitute already-known redefinitions. To use this option correctly, redefinitions must be expressed in terms of  \mmaInlineCell{Input}{MPhysSymbol[]}. In this context, the  \mmaInlineCell{Input}{ReplaceBareMass[]} function is particularly useful, as it automatically rewrites the input in terms of the symbolic physical mass. Imagine we know in advance the result from  \mmaInlineCell{Input}{Out[39]}, then we can perform the matching for the 6-$\phi$ interaction as:

\begin{mathematicaNotebook}
\begin{mmaCell}[moredefined={modelFull,sol1BareMass}]{Input}
  reduction = (#\partL1\partR\(\,\,\)->\(\,\,\)ReplaceBareMass[#\partL2\partR, 6, modelFull]) \& /@ sol1BareMass
\end{mmaCell}
\begin{mmaCell}{Output}
  \{\mmaSubSup{lmbd}{phys}{(4)}\(\,\,\)\(\to\)\(\,\,\)\mmaSubSup{lmbd}{full}{(4)},\(\,\,\)\mmaSubSup{lmbd}{phys}{(5)}\(\,\,\)\(\to\)\(\,\,\)0,
  
   \mmaSubSup{lmbd}{phys}{(6)}\(\,\,\)\(\to\)\(\,\,\)\mmaSup{MPhysSymbol[S[1]]}{2}(-8\(\,\,\)\mmaSubSup{lmbd}{full}{(4)} \mmaSubSup{rD\(\phi\)}{full}{(6)}+ \mmaSubSup{r\(\phi\)\,D}{full}{(6)})\}
  
\end{mmaCell}
\begin{mmaCell}[moredefined={reduction,modelFull,modelPhys}]{Input}
  sol2bis\(\,\,\)=\(\,\,\)AmplitudeMatching[modelFull,\(\,\,\)modelPhys,\(\,\,\)\{S[1],S[1],S[1],S[1],S[1],S[1]\},
  6,\(\,\,\)Replacements->\{\{\},reduction\}]
  
\end{mmaCell}
\begin{mmaCell}{Output}
  \{\mmaSubSup{a\(\phi\)}{phys}{(6)}\(\,\,\)\(\to\)\(\,\,\)\mmaSubSup{a\(\phi\)}{full}{(6)}+ 4\(\,\,\)\mmaSubSup{lmbd}{full}{(4)}(4\(\,\,\)\mmaSubSup{lmbd}{full}{(4)}\mmaSubSup{rD\(\phi\)}{full}{(6)}- \mmaSubSup{r\(\phi\)\,D}{full}{(6)})\}

\end{mmaCell}
\end{mathematicaNotebook}

This iterative, process-by-process substitution procedure constitutes the core workflow of the main function \mmaInlineCell{Input}{RedBasis[]}. Considering $\mathcal{L}_{full}$, the result for the reduction can be obtained by using this function as follows:
\begin{mathematicaNotebook}
\begin{mmaCell}[moredefined={modelFull,modelPhys}]{Input}
  finalReduction = RedBasis[modelFull, modelPhys, 6];
\end{mmaCell}
\begin{mmaCell}[moregraphics={moreig={scale=0.09},yoffset=0.5ex}]{Print}
  \mmaEcho{>> }Number of solved WCs: 5/5
  \mmaEcho{>> }List of solved WCs: \{m, g, lmbd, a\(\psi\), a\(\phi\)\}
  \mmaEcho{>> }Showing the reduction
  \mmaSub{m}{phys}\(\,\,\)\(\to\)\(\,\,\) \mmaSub{m}{full}- \mmaSubSup{m}{full}{3}\mmaSub{r\(\phi\)D}{full}

  \mmaSub{m\(\psi\)}{phys}\(\,\,\)\(\to\)\(\,\,\) \mmaSub{m\(\psi\)}{full}+ \mmaSubSup{m\(\psi\)}{full}{3}\mmaSub{r\(\psi\)D}{full}
  
  \mmaSub{lmbd}{phys}\(\,\,\)\(\to\)\(\,\,\)\mmaSub{lmbd}{full}- 8\(\,\,\)\mmaSub{lmbd}{full}\mmaSubSup{m}{full}{2}\mmaSub{rD\(\phi\)}{full}+ \mmaSubSup{m}{full}{2}\mmaSub{r\(\phi\)D}{full}
  
  \mmaSub{g}{phys}\(\,\,\)\(\to\)\(\,\,\) \mmaSub{g}{full}
  
  \mmaSub{a\(\phi\)}{phys}\(\,\,\)\(\to\)\(\,\,\)\mmaSub{a\(\phi\)}{full}+16\(\,\,\)\mmaSubSup{lmbd}{full}{2}\mmaSub{rD\(\phi\)}{full}- 4\(\,\,\)\mmaSub{lmbd}{full}\mmaSub{r\(\phi\)D}{full}
  
  \mmaSubSup{a\(\psi\)}{phys}{(6)}\(\,\,\)\(\to\)\(\,\,\)\mmaSub{a\(\psi\)}{full}- \mmaFrac{1}{2}\mmaSubSup{g}{full}{2}\mmaSub{r2F}{full} + \mmaSub{g}{full}\mmaSub{rDF\(\psi\)}{full}

  \mmaSubSup{{a\(\psi\)}F}{phys}{(6)}\(\,\,\)\(\to\)\(\,\,\)\mmaSub{{a\(\psi\)}F}{full}- \mmaFrac{1}{2}\mmaSub{g}{full}\mmaSub{m\(\psi\)}{full} \mmaSub{r\(\psi\)D}{full} 

  All done! \mmaGraphics{Images/mosquita}

  \mmaEcho{>> Elapsed time:} 3.594948
  
\end{mmaCell}
\end{mathematicaNotebook}

\subsection{Useful functions}

\label{sect: Usefil Functions}

This section is dedicated to a set of auxiliary functions that, while not directly involved in the reduction of a Lagrangian, play a supportive role in the main workflow of \tool. These can be useful for users to carry out algebraic transformations, expansions, and other symbolic manipulations on expressions, providing greater flexibility and control in intermediate steps.

The first function, used inside most \tool\ functions, is 
\begin{functiondef}
\mmaInlineCell[moredefined=expression]{Input}{EFTSeries[expression, EFTOrder]}   
\end{functiondef}
\noindent
which performs a series expansion in the perturbative parameter \mmaInlineCell{Input}{inv\(\Lambda\)} to a given EFT dimension, that is, to order \mmaInlineCell{Input}{\mmaSup{inv\(\Lambda\)}{(EFTOrder-4)}} \footnote{If the EFT order is set to an integer lower than 4, it computes the ``renormalizable'' (or EFT-independent) part of the expression.}. It is important to notice that, by default, it applies an
\begin{functiondef}
\mmaInlineCell[moredefined=expression]{Input}{ExplicitEFTOrder[expression]}
\end{functiondef}
\noindent
before doing the series expansion. This last function, which we encountered before, first ignores every occurrence of \mmaInlineCell{Input}{inv\(\Lambda\)} in \mmaInlineCell[moredefined=expression]{Input}{expression} and then gets the right power of the EFT parameter according to the products of different WCs and their EFT order, independently of the model they belong to. For instance, we would have the following behaviors:

\begin{mathematicaNotebook}
\begin{mmaCell}[moredefined={model1,model2}]{Input}
  ExplicitEFTOrder[inv\(\Lambda\)^2 \(\alpha\)]
  ExplicitEFTOrder[inv\(\Lambda\)^2 \(\alpha\) WC[c1, 6, model1] WC[c2, 6, model2]]
  
\end{mmaCell}
\begin{mmaCell}{Output}
  \(\alpha\)
  \mmaSup{inv\(\Lambda\)}{4}\(\,\,\)\(\alpha\)\(\,\,\)\mmaSubSup{c1}{model1}{(6)}\(\,\,\)\mmaSubSup{c2}{model2}{(6)}
  
\end{mmaCell}
\end{mathematicaNotebook}

\noindent
If one wants to compute a series in the EFT parameter, explicitly leaving the powers of \mmaInlineCell{Input}{inv\(\Lambda\)} already present in \mmaInlineCell[moredefined=expression]{Input}{expression}, it is possible to skip the \mmaInlineCell{Input}{ExplicitEFTOrder[]} command by setting \mmaInlineCell{Input}{ExplicitEFTOrder\(\,\,\)->\(\,\,\)False} in the \mmaInlineCell{Input}{EFTSeries}.

Finally, the matching equations \mmaInlineCell[moredefined=eqs]{Input}{eqs} obtained after replacing numerical kinematics are solved by means of the 
\begin{functiondef}
\mmaInlineCell[moredefined={eqs,solveVars}]{Input}{SolvePerturbative[eqs, solveVars]}
\end{functiondef}
\noindent
command, which automates the process of solving the system of equations \mmaInlineCell[moredefined=eqs]{Input}{eqs} in the variables \mmaInlineCell[moredefined=solveVars]{Input}{solveVars} order by order in \mmaInlineCell{Input}{inv\(\Lambda\)}. It solves the system by identifying and matching the coefficients of the different powers of \mmaInlineCell{Input}{inv\(\Lambda\)}, in a strictly increasing order and replacing the results of lower powers in the following matching equations. This significantly simplifies the system, because WCs appearing in products in higher order equations, \textit{e.g.} \mmaInlineCell[moredefined={c1,c2,m}]{Input}{\mmaSup{c1}{(6)}\mmaSup{c2}{(6)}}, could have already been solved in more simple linear equations at lower orders. Indeed, this function intends to solve the system of equations in a linear way, unless a non-linear equation is encountered, in which case it raises an error but continues with the resolution of the system with the \mathematica\ built-in function \mmaInlineCell{Input}{Solve}. This procedure ensures the finding of a solution, while a brute-force \mmaInlineCell[moredefined={eqs,solveVars}]{Input}{Solve[eqs,solveVars]} approach could fail in systems of really long equations, taking for hours and finding no solution, even when a solution does exist.

\section{Using \tool}
\label{sect:using mosca}

While the reader is already equipped with all the necessary tools to run and use \tool, we would like to devote this section to comment on the capabilities of \tool\ and show some possible issues and other aspects.

First, despite \tool\ being designed to facilitate the reduction of bases and implement the on-shell matching approach, we had to include a couple of functionalities of a much broader usage. Indeed, the automation of finding Feynman diagrams related by isomorphisms in order to reduce as much as possible the number of diagrams to compute is completely suitable for the everyday use of \texttt{FeynArts}, as far as all particles are considered incoming. This can be really convenient for both memory and time efficiency, as well as for facilitating several analysis aspects. Certainly, regarding this last statement, in the inspection of the different contributions to a given amplitude, it is much easier to visualize and understand unique Feynman diagrams (in the sense that they are not related by permutations) than to face all the---possibly hundreds of---graphs returned by \texttt{FeynArts}. 

\begin{figure}[ht]
    \centering
    \includegraphics[width=\linewidth]{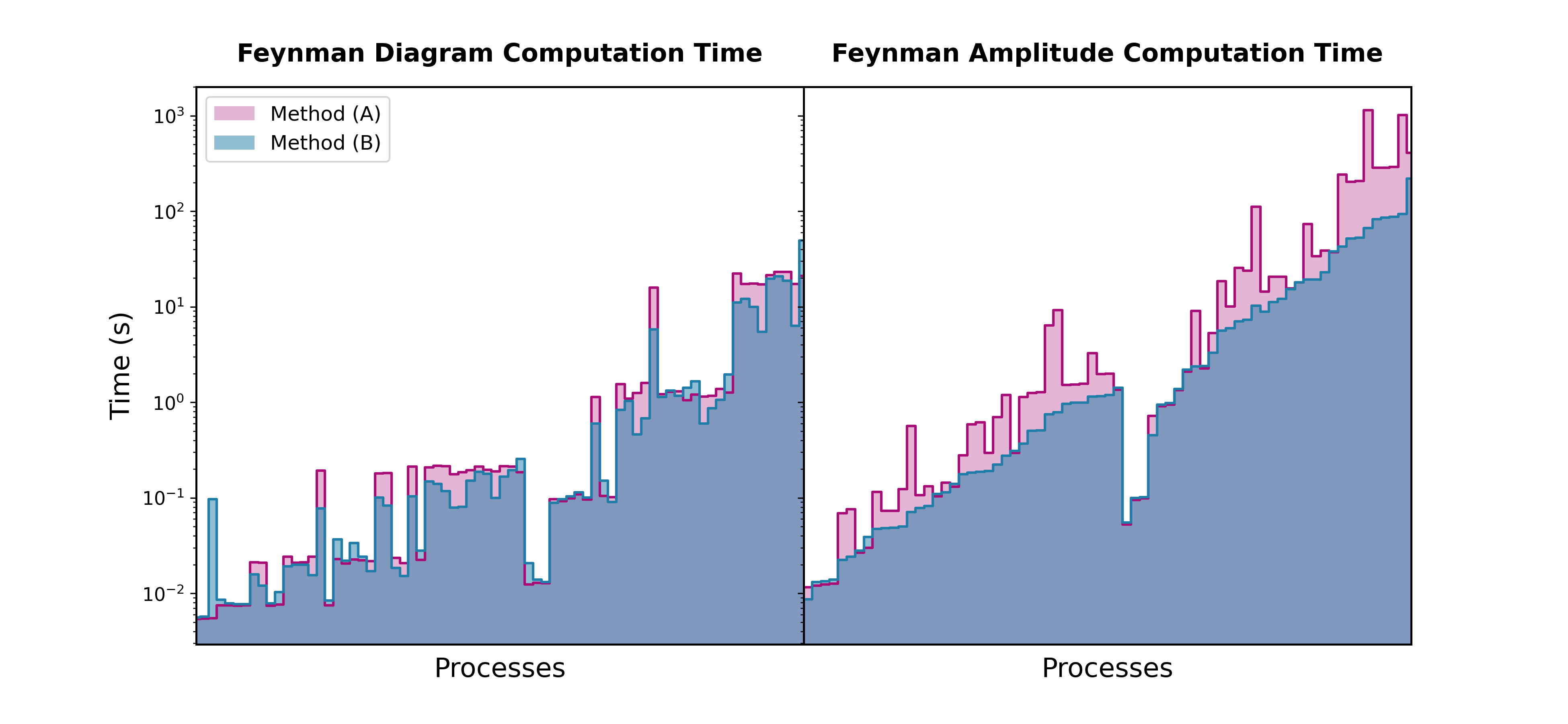}
    \caption{Comparison of diagram and amplitude computation times for 76 different processes within the EW sector of dimension-6 SMEFT Green's basis with two different methods: standard \texttt{FeynArts} techniques (A) and our isomorphisms-identification approach (B). The amplitude computation time in method (B) corresponds to the total time after applying a \texttt{RecoverFullAmplitude[]} command.}
    \label{fig:diagram and amplitude time}
\end{figure}

As far as time and memory efficiency are considered, we can take the electroweak (EW) sector of the SMEFT Green's basis for dimension 6 (\texttt{model\_SMEFT\_HBW}, provided within \tool) and compare some results if we use standard \texttt{FeynArts} techniques (A) or our isomorphisms-identification approach (B). In Figure \ref{fig:diagram and amplitude time} we show the comparison between methods in the computation time for diagrams and amplitude expressions in 76 different processes (the first 40 at tree-level and the following 36 at one-loop level).

Regarding the time it takes to just compute diagrams we can see a similar behavior, with slight performance differences depending on the process. However, a notable improvement of method (B) with respect to method (A) is found after using the \mmaInlineCell{Input}{FCFAConvert[CreateFeynAmp[#]]} command to compute the associated amplitude. This obviously extrapolates to every subsequent manipulation performed over the amplitudes, saving considerable time and memory if those are only applied to the minimal set of permutation-unrelated diagrams (and afterwards reconstructing the whole amplitude with the permutations, if needed), instead of applying them directly over all diagrams.

Note that the $Y$-axis in the graphs is set to logarithmic scale. Thus, there are some processes for which the amplitude computation time varies from 1 to 30 minutes, depending on the employed method. Moreover, the reason why there are only 36 one-loop processes instead of 40 is that the full amplitude expressions for the last 4 chosen processes were not even manageable by the available computer memory. With approach (B) this does not represent an issue, as long as one works with the minimal set of permutation-unrelated amplitudes.

Other general-purpose tools than can be used beyond on-shell matching context are the commands to perform rational kinematic substitutions. Although there are some package implementations to generate random-valued kinematics (see refs. \cite{AccettulliHuber:2023ldr,Maitre:2007jq}), to the best of our knowledge, there is no such tool to explicitly replace rational-valued kinematics satisfying on-shell conditions in analytic Feynman amplitudes, with even the possibility to keep masses symbolic. Indeed, although the \mmaInlineCell{Input}{ReplaceKinematics[]} is absolutely designed to translate expressions from \texttt{FeynCalc}, the \mmaInlineCell{Input}{KinematicConfigurations[]} function is completely general, so that a generalization of \mmaInlineCell{Input}{ReplaceKinematics[]} to interpret expressions from other packages, whether private or public, such as \texttt{FeynArts} itself or \texttt{FormCalc} \cite{Hahn:1998yk}, should be rather straightforward.

After these comments on general uses of different commands, we would like to give some advice when using \tool. First, performing a \mmaInlineCell{Input}{FormatWC[]} is almost mandatory if one does not want to decipher unintelligible outputs in terms of \mmaInlineCell{Input}{WC[]} objects. Following this, while setting \mmaInlineCell{Input}{VisibleOrder} and \mmaInlineCell{Input}{VisibleModel} options to \mmaInlineCell{Input}{True} can blur things a bit, they can be very useful to check, for example, whether a WC has been completely reduced or not after an \mmaInlineCell{Input}{AmplitudeMatching[]}. Indeed, it could happen than we enocunter an output like \mmaInlineCell[moredefined={c1,c2,r}]{Input}{c1->c1\(\,\,\)+\(\,\,\)r\(\,\)c2}, coming from a command similar to  \mmaInlineCell[moredefined={model1,model2,process,X}]{Input}{AmplitudeMatching[model1,model2,process,X]}. In light of this, it may seem that the WC $c_1$ has been matched. But then, after a \mmaInlineCell{Input}{VisibleModel->True} setting, it could turn out that \mmaInlineCell[moredefined={c1,c2,r,model1,model2}]{Input}{\mmaSub{c1}{model2}->\mmaSub{c1}{model1}\(\,\,\)+\(\,\,\)\mmaSub{r}{model1}\(\,\)\mmaSub{c2}{model2}}, revealing that the matching procedure was not able to disentangle $c_2$ from $c_1$. In such a case, it would yet be necessary another process to fully have the expressions for $c_1$ and $c_2$ only in terms of WCs from  \mmaInlineCell[moredefined={model1}]{Input}{model1}. 

Another useful tip is regarding the \mmaInlineCell{Input}{RedBasis[]} function. This command does not stop until it either matches every WC in the model or computes every available process in \mmaInlineCell{Input}{ProcessList[]}. Yet, and despite all efforts to speed up computations, for amplitudes with a high number of external legs the procedure is sometimes remarkably slow. When facing one of those processes, one could be tempted to abort the evaluation, at the cost of losing the collected solutions of all already-matched WCs. To avoid this, every call to \mmaInlineCell[moredefined={model1,model2,X}]{Input}{RedBasis[model1,model2,X]} internally appends the solution of the matching after every process to a list, under the name of the reserved keyword \mmaInlineCell[moredefined={model1,model2,X}]{Input}{Reduction[model1,model2,X]}. Hence, if one aborts the evaluation, there is no information lost.

Bear in mind that \mmaInlineCell[moredefined={model1,model2,X}]{Input}{Reduction[model1,model2,X]} essentially stores the results given by a series of \mmaInlineCell{Input}{AmplitudeMatching[]} calls, so that: (1) they are returned in the order-by-order expansion of WCs, (2) they are written in terms of \mmaInlineCell{Input}{MPhysSymbol[]} and (3) it could happen that ${c_k}_\mathrm{model2} \to f({c_i}_\mathrm{model1}, {c_j}_\mathrm{model2})$ where $f$ is a function of WCs from both \mmaInlineCell[moredefined=model1]{Input}{model1} and \mmaInlineCell[moredefined=model2]{Input}{model2}, as noted previously. For (1) and (2), there exist built-in functions such as \mmaInlineCell{Input}{JoinOrderWC[]} and \mmaInlineCell{Input}{ReplacePhysMass[]} to deal with these ``inconveniences''. For (3), it could be the case that some of the ${c_j}_\mathrm{model2}$ appearing in the redefinition of ${c_k}_\mathrm{model2}$ have been solved later on; hence, it is necessary to perform a repeated replacement of the list of reductions over itself (of course, only over the values, not the keys). In this vein, we also would like to recommend the users that, in case they want to save results to a text file, do so with a \mmaInlineCell[pattern={c_,c}]{Input}{/.WC[c_,__]:>c} replacement rule, in order to avoid the heavily formatted \mmaInlineCell{Input}{WC} structures.

Finally, the \tool\ folder includes a collection of example notebooks (\texttt{mosca-notebook-X.nb}), which demonstrate the usage of the package. The models used in these examples are stored in the \texttt{Models/} directory. Additionally, the \texttt{mosca-paper-example.nb} notebook is provided, allowing you to try out some of the commands discussed throughout this paper.

\section{Conclusions and outlook}
\label{sect: conclusions}

In this article, we have introduced the first version of \tool, which implements an efficient on-shell matching algorithm. The toolkit includes several additional features designed to facilitate the manipulation of symbolic expressions. Among these, there are functions that implement isomorphism-identification algorithms to simplify the computation of Feynman diagrams, as well as utilities for performing algebraic transformations, series expansions, and other symbolic manipulations. As demonstrated in Section \ref{sect:using mosca}, the automation of identifying isomorphic Feynman diagrams results in a non-negligible reduction in computation time. Furthermore, the main function of the package in this initial release is dedicated to automatically perform reduction of redundant bases to physical bases in effective field theories, via the on-shell matching procedure.

As practical examples we have shown how to use \tool\ in several different situations, including: the calculation of physical masses, wavefunction factors and explicit forms of Feynman propagators; the computation of amplitudes for distinct processes involving scalar, fermion and vector fields, with the corresponding identification of isomorphic diagrams and substitution of kinematic structures; and finally, the reduction of a Green's basis to a physical one by means of the main function of \tool.

Looking ahead, there are several directions in which \tool\ can naturally evolve. First, in this initial release the package cannot deal with flavor indices, so we will make it compatible. Second, we will also consider the possibility of having 2-point operators with mixing fields. Finally, from a broader perspective, we plan to extend the capabilities of \tool\ to the loop level, which includes the direct renormalization of effective Lagrangians in a physical basis, as well as the automated computation of finite matching contributions, including those from evanescent operators. These extensions have been previously discussed in \cite{chala2024efficientonshellmatching}, which demonstrates the effectiveness of numerical on-shell methods even in these scenarios. While implementing these features is expected to be relatively straightforward in principle, special care must be taken in treating the soft contributions to amplitudes in the full and effective theories, as they might be different due to evanescent effects.

\tool\ represents a significant development in the emerging field of on-shell methods, a framework that remains far less developed and barely adopted compared to the traditional off-shell approach. Once all planned functionalities are implemented, \tool\ will fully support a complete on-shell matching procedure based solely on a physical operator basis. This eliminates the need to include redundant or evanescent operators and avoids the often cumbersome process of basis reduction. 

\section*{Acknowledgments}

We would like to acknowledge the valuable contributions of Mikael Chala and José Santiago during the development of the package. As \tool\ is the result of a previously shared collaboration, the insightful discussions and ideas exchanged with them were essential to its success. We also acknowledge support from the MCIN/AEI (10.13039/501100011033) and ERDF (grants PID2022-139466NB-C21 and PID2022-139466NB-C22), from the Junta de Andalucía grant P21-00199 and from Consejería de Universidad, Investigación e Innovación, Gobierno de España and Unión Europea -- NextGenerationEU under grants AST22 6.5 and CNS2022-136024. JLM is further supported by an FPU grant (FPU23/02028) from Consejería de Universidad, Investigación e Innovación, Gobierno de España.

\setcounter{biburlnumpenalty}{7000}
\setcounter{biburllcpenalty}{7000}
\setcounter{biburlucpenalty}{7000}



\renewcommand{\bibfont}{\small}
\phantomsection
\printbibheading[title={References}, heading = subbibintoc]
\printbibliography[heading=none]

@article{Criado:2018sdb,
    author = "Criado, J. C. and P\'erez-Victoria, M.",
    title = "{Field redefinitions in effective theories at higher orders}",
    eprint = "1811.09413",
    archivePrefix = "arXiv",
    primaryClass = "hep-ph",
    doi = "10.1007/JHEP03(2019)038",
    journal = "JHEP",
    volume = "03",
    pages = "038",
    year = "2019"
}

@article{Arzt:1993gz,
    author = "Arzt, Christopher",
    title = "{Reduced effective Lagrangians}",
    eprint = "hep-ph/9304230",
    archivePrefix = "arXiv",
    reportNumber = "UM-TH-92-28",
    doi = "10.1016/0370-2693(94)01419-D",
    journal = "Phys. Lett. B",
    volume = "342",
    pages = "189--195",
    year = "1995"
}

@article{Fuentes-Martin:2022jrf,
    author = {Fuentes-Mart\'\i{}n, Javier and K\"onig, Matthias and Pag\`es, Julie and Thomsen, Anders Eller and Wilsch, Felix},
    title = "{A proof of concept for matchete: an automated tool for matching effective theories}",
    eprint = "2212.04510",
    archivePrefix = "arXiv",
    primaryClass = "hep-ph",
    reportNumber = "MITP-22-105, TUM-HEP-1443/22, ZU-TH-58/22",
    doi = "10.1140/epjc/s10052-023-11726-1",
    journal = "Eur. Phys. J. C",
    volume = "83",
    number = "7",
    pages = "662",
    year = "2023"
}

@article{Carmona:2021xtq,
    author = "Carmona, Adrian and Lazopoulos, Achilleas and Olgoso, Pablo and Santiago, Jose",
    title = "{Matchmakereft: automated tree-level and one-loop matching}",
    eprint = "2112.10787",
    archivePrefix = "arXiv",
    primaryClass = "hep-ph",
    doi = "10.21468/SciPostPhys.12.6.198",
    journal = "SciPost Phys.",
    volume = "12",
    number = "6",
    pages = "198",
    year = "2022"
}

@misc{chala2024efficientonshellmatching,
      title={Efficient on-shell matching}, 
      author={Mikael Chala and Javier López Miras and José Santiago and Fuensanta Vilches},
      year={2024},
      eprint={2411.12798},
      archivePrefix={arXiv},
      primaryClass={hep-ph},
      url={https://arxiv.org/abs/2411.12798} 
}

@article{Brambilla:2020fla,
    author = "Brambilla, Nora and Chung, Hee Sok and Shtabovenko, Vladyslav and Vairo, Antonio",
    title = "{FeynOnium: Using FeynCalc for automatic calculations in Nonrelativistic Effective Field Theories}",
    eprint = "2006.15451",
    archivePrefix = "arXiv",
    primaryClass = "hep-ph",
    reportNumber = "TUM-EFT 75/15, TTP19-021",
    doi = "10.1007/JHEP11(2020)130",
    journal = "JHEP",
    volume = "11",
    pages = "130",
    year = "2020"
}

@article{Hahn:2000kx,
    author = "Hahn, Thomas",
    title = "{Generating Feynman diagrams and amplitudes with FeynArts 3}",
    eprint = "hep-ph/0012260",
    archivePrefix = "arXiv",
    reportNumber = "KA-TP-23-2000",
    doi = "10.1016/S0010-4655(01)00290-9",
    journal = "Comput. Phys. Commun.",
    volume = "140",
    pages = "418--431",
    year = "2001"
}

@article{Hodges:2009hk,
    author = "Hodges, Andrew",
    title = "{Eliminating spurious poles from gauge-theoretic amplitudes}",
    eprint = "0905.1473",
    archivePrefix = "arXiv",
    primaryClass = "hep-th",
    doi = "10.1007/JHEP05(2013)135",
    journal = "JHEP",
    volume = "05",
    pages = "135",
    year = "2013"
}

@article{DeAngelis:2022qco,
    author = "De Angelis, Stefano",
    title = "{Amplitude bases in generic EFTs}",
    eprint = "2202.02681",
    archivePrefix = "arXiv",
    primaryClass = "hep-th",
    reportNumber = "QMUL-PH-22-05, SAGEX-22-18-E",
    doi = "10.1007/JHEP08(2022)299",
    journal = "JHEP",
    volume = "08",
    pages = "299",
    year = "2022"
}

@article{Kluberg-Stern:1975ebk,
    author = "Kluberg-Stern, H. and Zuber, J. B.",
    title = "{Renormalization of Nonabelian Gauge Theories in a Background Field Gauge. 2. Gauge Invariant Operators}",
    reportNumber = "SACLAY-DPh-T/75/28",
    doi = "10.1103/PhysRevD.12.3159",
    journal = "Phys. Rev. D",
    volume = "12",
    pages = "3159--3180",
    year = "1975"
}

@article{Grosse-Knetter:1993tae,
    author = "Grosse-Knetter, Carsten",
    title = "{Effective Lagrangians with higher derivatives and equations of motion}",
    eprint = "hep-ph/9306321",
    archivePrefix = "arXiv",
    reportNumber = "BI-TP-93-29",
    doi = "10.1103/PhysRevD.49.6709",
    journal = "Phys. Rev. D",
    volume = "49",
    pages = "6709--6719",
    year = "1994"
}

@article{Wudka:1994ny,
    author = "Wudka, Jose",
    title = "{Electroweak effective Lagrangians}",
    eprint = "hep-ph/9406205",
    archivePrefix = "arXiv",
    reportNumber = "UCRHEP-T-121A, UCRHEP-T121",
    doi = "10.1142/S0217751X94000959",
    journal = "Int. J. Mod. Phys. A",
    volume = "9",
    pages = "2301--2362",
    year = "1994"
}

@article{Manohar:2018aog,
    author = "Manohar, Aneesh V.",
    editor = "Davidson, Sacha and Gambino, Paolo and Laine, Mikko and Neubert, Matthias and Salomon, Christophe",
    title = "{Introduction to Effective Field Theories}",
    eprint = "1804.05863",
    archivePrefix = "arXiv",
    primaryClass = "hep-ph",
    doi = "10.1093/oso/9780198855743.003.0002",
    month = "4",
    year = "2018"
}

@article{AccettulliHuber:2023ldr,
    author = "Accettulli Huber, Manuel",
    title = "{SpinorHelicity4D: a Mathematica toolbox for the four-dimensional spinor-helicity formalism}",
    eprint = "2304.01589",
    archivePrefix = "arXiv",
    primaryClass = "hep-th",
    month = "4",
    year = "2023"
}

@article{Arkani-Hamed:2017jhn,
    author = "Arkani-Hamed, Nima and Huang, Tzu-Chen and Huang, Yu-tin",
    title = "{Scattering amplitudes for all masses and spins}",
    eprint = "1709.04891",
    archivePrefix = "arXiv",
    primaryClass = "hep-th",
    reportNumber = "NCTS-TH/1714, NCTS-TH-1714",
    doi = "10.1007/JHEP11(2021)070",
    journal = "JHEP",
    volume = "11",
    pages = "070",
    year = "2021"
}

@article{Banerjee:2022thk,
    author = "Banerjee, Upalaparna and Chakrabortty, Joydeep and Englert, Christoph and Rahaman, Shakeel Ur and Spannowsky, Michael",
    title = "{Integrating out heavy scalars with modified equations of motion: Matching computation of dimension-eight SMEFT coefficients}",
    eprint = "2210.14761",
    archivePrefix = "arXiv",
    primaryClass = "hep-ph",
    doi = "10.1103/PhysRevD.107.055007",
    journal = "Phys. Rev. D",
    volume = "107",
    number = "5",
    pages = "055007",
    year = "2023"
}

@article{Cheung:2009dc,
    author = "Cheung, Clifford and O'Connell, Donal",
    title = "{Amplitudes and Spinor-Helicity in Six Dimensions}",
    eprint = "0902.0981",
    archivePrefix = "arXiv",
    primaryClass = "hep-th",
    doi = "10.1088/1126-6708/2009/07/075",
    journal = "JHEP",
    volume = "07",
    pages = "075",
    year = "2009"
}

@article{Maitre:2007jq,
    author = "Maitre, D. and Mastrolia, P.",
    title = "{S@M, a Mathematica Implementation of the Spinor-Helicity Formalism}",
    eprint = "0710.5559",
    archivePrefix = "arXiv",
    primaryClass = "hep-ph",
    reportNumber = "SLAC-PUB-12867, ZU-TH-25-07",
    doi = "10.1016/j.cpc.2008.05.002",
    journal = "Comput. Phys. Commun.",
    volume = "179",
    pages = "501--574",
    year = "2008"
}

@article{Hahn:1998yk,
    author = "Hahn, T. and Perez-Victoria, M.",
    title = "{Automatized one loop calculations in four-dimensions and D-dimensions}",
    eprint = "hep-ph/9807565",
    archivePrefix = "arXiv",
    reportNumber = "UG-FT-87-98, KA-TP-7-1998",
    doi = "10.1016/S0010-4655(98)00173-8",
    journal = "Comput. Phys. Commun.",
    volume = "118",
    pages = "153--165",
    year = "1999"
}

@article{Hartland:2019bjb,
    author = "Hartland, Nathan P. and Maltoni, Fabio and Nocera, Emanuele R. and Rojo, Juan and Slade, Emma and Vryonidou, Eleni and Zhang, Cen",
    title = "{A Monte Carlo global analysis of the Standard Model Effective Field Theory: the top quark sector}",
    eprint = "1901.05965",
    archivePrefix = "arXiv",
    primaryClass = "hep-ph",
    reportNumber = "OUTP-18-07P, Nikhef-2018-058, CP3-19-02, CERN-TH-2018-274",
    doi = "10.1007/JHEP04(2019)100",
    journal = "JHEP",
    volume = "04",
    pages = "100",
    year = "2019"
}

@article{Aoude:2020dwv,
    author = "Aoude, Rafael and Hurth, Tobias and Renner, Sophie and Shepherd, William",
    title = "{The impact of flavour data on global fits of the MFV SMEFT}",
    eprint = "2003.05432",
    archivePrefix = "arXiv",
    primaryClass = "hep-ph",
    doi = "10.1007/JHEP12(2020)113",
    journal = "JHEP",
    volume = "12",
    pages = "113",
    year = "2020"
}

@article{Dawson:2020oco,
    author = "Dawson, Sally and Homiller, Samuel and Lane, Samuel D.",
    title = "{Putting standard model EFT fits to work}",
    eprint = "2007.01296",
    archivePrefix = "arXiv",
    primaryClass = "hep-ph",
    reportNumber = "YITP-SB-20-18",
    doi = "10.1103/PhysRevD.102.055012",
    journal = "Phys. Rev. D",
    volume = "102",
    number = "5",
    pages = "055012",
    year = "2020"
}

@article{Anisha:2020ggj,
    author = "Anisha and Das Bakshi, Supratim and Chakrabortty, Joydeep and Patra, Sunando Kumar",
    title = "{Connecting electroweak-scale observables to BSM physics through EFT and Bayesian statistics}",
    eprint = "2010.04088",
    archivePrefix = "arXiv",
    primaryClass = "hep-ph",
    doi = "10.1103/PhysRevD.103.076007",
    journal = "Phys. Rev. D",
    volume = "103",
    number = "7",
    pages = "076007",
    year = "2021"
}

@article{Falkowski:2020pma,
    author = "Falkowski, Adam and Gonz\'alez-Alonso, Mart\'\i{}n and Naviliat-Cuncic, Oscar",
    title = "{Comprehensive analysis of beta decays within and beyond the Standard Model}",
    eprint = "2010.13797",
    archivePrefix = "arXiv",
    primaryClass = "hep-ph",
    reportNumber = "IFIC/20-49, FTUV/20-1027",
    doi = "10.1007/JHEP04(2021)126",
    journal = "JHEP",
    volume = "04",
    pages = "126",
    year = "2021"
}

@article{Ellis:2020unq,
    author = "Ellis, John and Madigan, Maeve and Mimasu, Ken and Sanz, Veronica and You, Tevong",
    title = "{Top, Higgs, Diboson and Electroweak Fit to the Standard Model Effective Field Theory}",
    eprint = "2012.02779",
    archivePrefix = "arXiv",
    primaryClass = "hep-ph",
    reportNumber = "KCL-PH-TH/2020-73, CERN-TH-2020-202",
    doi = "10.1007/JHEP04(2021)279",
    journal = "JHEP",
    volume = "04",
    pages = "279",
    year = "2021"
}

@article{Ethier:2021bye,
    author = "Ethier, Jacob J. and Magni, Giacomo and Maltoni, Fabio and Mantani, Luca and Nocera, Emanuele R. and Rojo, Juan and Slade, Emma and Vryonidou, Eleni and Zhang, Cen",
    collaboration = "SMEFiT",
    title = "{Combined SMEFT interpretation of Higgs, diboson, and top quark data from the LHC}",
    eprint = "2105.00006",
    archivePrefix = "arXiv",
    primaryClass = "hep-ph",
    reportNumber = "OUTP-20-05P, Nikhef-2020-020, CP3-21-12, MCNET-21-07,
  MAN/HEP/2021/004",
    doi = "10.1007/JHEP11(2021)089",
    journal = "JHEP",
    volume = "11",
    pages = "089",
    year = "2021"
}

@inproceedings{deBlas:2022ofj,
    author = "de Blas, Jorge and Du, Yong and Grojean, Christophe and Gu, Jiayin and Miralles, Victor and Peskin, Michael E. and Tian, Junping and Vos, Marcel and Vryonidou, Eleni",
    title = "{Global SMEFT Fits at Future Colliders}",
    booktitle = "{Snowmass 2021}",
    eprint = "2206.08326",
    archivePrefix = "arXiv",
    primaryClass = "hep-ph",
    month = "6",
    year = "2022"
}

@article{Grzadkowski:2010es,
    author = "Grzadkowski, B. and Iskrzynski, M. and Misiak, M. and Rosiek, J.",
    title = "{Dimension-Six Terms in the Standard Model Lagrangian}",
    eprint = "1008.4884",
    archivePrefix = "arXiv",
    primaryClass = "hep-ph",
    reportNumber = "IFT-9-2010, TTP10-35",
    doi = "10.1007/JHEP10(2010)085",
    journal = "JHEP",
    volume = "10",
    pages = "085",
    year = "2010"
}

@article{delAguila:2008ir,
    author = "del Aguila, Francisco and Bar-Shalom, Shaouly and Soni, Amarjit and Wudka, Jose",
    title = "{Heavy Majorana Neutrinos in the Effective Lagrangian Description: Application to Hadron Colliders}",
    eprint = "0806.0876",
    archivePrefix = "arXiv",
    primaryClass = "hep-ph",
    reportNumber = "UG-FT-230-08, CAFPE-100-08, UCI-TR-2008-21, BNL-HET-08-14",
    doi = "10.1016/j.physletb.2008.11.031",
    journal = "Phys. Lett. B",
    volume = "670",
    pages = "399--402",
    year = "2009"
}

@article{Aparici:2009fh,
    author = "Aparici, Alberto and Kim, Kyungwook and Santamaria, Arcadi and Wudka, Jose",
    title = "{Right-handed neutrino magnetic moments}",
    eprint = "0904.3244",
    archivePrefix = "arXiv",
    primaryClass = "hep-ph",
    reportNumber = "FTUV-09-0421, IFIC-09-15, UCRHEP-T466",
    doi = "10.1103/PhysRevD.80.013010",
    journal = "Phys. Rev. D",
    volume = "80",
    pages = "013010",
    year = "2009"
}

@article{Bhattacharya:2015vja,
    author = "Bhattacharya, Subhaditya and Wudka, Jos\'e",
    title = "{Dimension-seven operators in the standard model with right handed neutrinos}",
    eprint = "1505.05264",
    archivePrefix = "arXiv",
    primaryClass = "hep-ph",
    doi = "10.1103/PhysRevD.94.055022",
    journal = "Phys. Rev. D",
    volume = "94",
    number = "5",
    pages = "055022",
    year = "2016",
    note = "[Erratum: Phys.Rev.D 95, 039904 (2017)]"
}

@article{Liao:2016qyd,
    author = "Liao, Yi and Ma, Xiao-Dong",
    title = "{Operators up to Dimension Seven in Standard Model Effective Field Theory Extended with Sterile Neutrinos}",
    eprint = "1612.04527",
    archivePrefix = "arXiv",
    primaryClass = "hep-ph",
    doi = "10.1103/PhysRevD.96.015012",
    journal = "Phys. Rev. D",
    volume = "96",
    number = "1",
    pages = "015012",
    year = "2017"
}

@article{Aebischer:2024csk,
    author = "Aebischer, Jason and Kapoor, Tejhas and Kumar, Jacky",
    title = "{wilson: A package for renormalization group running in the SMEFT with Sterile Neutrinos}",
    eprint = "2411.07220",
    archivePrefix = "arXiv",
    primaryClass = "hep-ph",
    reportNumber = "CERN-TH-2024-186, LA-UR-24-31845",
    month = "11",
    year = "2024"
}

@article{Jenkins:2017jig,
    author = "Jenkins, Elizabeth E. and Manohar, Aneesh V. and Stoffer, Peter",
    title = "{Low-Energy Effective Field Theory below the Electroweak Scale: Operators and Matching}",
    eprint = "1709.04486",
    archivePrefix = "arXiv",
    primaryClass = "hep-ph",
    doi = "10.1007/JHEP03(2018)016",
    journal = "JHEP",
    volume = "03",
    pages = "016",
    year = "2018"
}

@article{Bischer:2019ttk,
    author = "Bischer, Ingolf and Rodejohann, Werner",
    title = "{General neutrino interactions from an effective field theory perspective}",
    eprint = "1905.08699",
    archivePrefix = "arXiv",
    primaryClass = "hep-ph",
    doi = "10.1016/j.nuclphysb.2019.114746",
    journal = "Nucl. Phys. B",
    volume = "947",
    pages = "114746",
    year = "2019"
}

@article{Li:2020lba,
    author = "Li, Tong and Ma, Xiao-Dong and Schmidt, Michael A.",
    title = "{General neutrino interactions with sterile neutrinos in light of coherent neutrino-nucleus scattering and meson invisible decays}",
    eprint = "2005.01543",
    archivePrefix = "arXiv",
    primaryClass = "hep-ph",
    doi = "10.1007/JHEP07(2020)152",
    journal = "JHEP",
    volume = "07",
    pages = "152",
    year = "2020"
}

@article{Li:2020wxi,
    author = "Li, Tong and Ma, Xiao-Dong and Schmidt, Michael A.",
    title = "{Constraints on the charged currents in general neutrino interactions with sterile neutrinos}",
    eprint = "2007.15408",
    archivePrefix = "arXiv",
    primaryClass = "hep-ph",
    doi = "10.1007/JHEP10(2020)115",
    journal = "JHEP",
    volume = "10",
    pages = "115",
    year = "2020"
}

@article{Chala:2020vqp,
    author = "Chala, Mikael and Titov, Arsenii",
    title = "{One-loop matching in the SMEFT extended with a sterile neutrino}",
    eprint = "2001.07732",
    archivePrefix = "arXiv",
    primaryClass = "hep-ph",
    doi = "10.1007/JHEP05(2020)139",
    journal = "JHEP",
    volume = "05",
    pages = "139",
    year = "2020"
}

@article{Bauer:2020jbp,
    author = "Bauer, Martin and Neubert, Matthias and Renner, Sophie and Schnubel, Marvin and Thamm, Andrea",
    title = "{The Low-Energy Effective Theory of Axions and ALPs}",
    eprint = "2012.12272",
    archivePrefix = "arXiv",
    primaryClass = "hep-ph",
    reportNumber = "IPPP/20/69, MITP/20-070 SISSA 30/2020/FISI, ZH-TH-47/20",
    doi = "10.1007/JHEP04(2021)063",
    journal = "JHEP",
    volume = "04",
    pages = "063",
    year = "2021"
}

@article{Gripaios:2016xuo,
    author = "Gripaios, Ben and Sutherland, Dave",
    title = "{An operator basis for the Standard Model with an added scalar singlet}",
    eprint = "1604.07365",
    archivePrefix = "arXiv",
    primaryClass = "hep-ph",
    reportNumber = "CAVENDISH-HEP-16-05, NSF-KITP-16-056",
    doi = "10.1007/JHEP08(2016)103",
    journal = "JHEP",
    volume = "08",
    pages = "103",
    year = "2016"
}

@article{Grojean:2023tsd,
    author = "Grojean, Christophe and Kley, Jonathan and Yao, Chang-Yuan",
    title = "{Hilbert series for ALP EFTs}",
    eprint = "2307.08563",
    archivePrefix = "arXiv",
    primaryClass = "hep-ph",
    reportNumber = "DESY-23-098, HU-EP-23/39",
    doi = "10.1007/JHEP11(2023)196",
    journal = "JHEP",
    volume = "11",
    pages = "196",
    year = "2023"
}

@article{Chala:2020wvs,
    author = "Chala, Mikael and Guedes, Guilherme and Ramos, Maria and Santiago, Jose",
    title = "{Running in the ALPs}",
    eprint = "2012.09017",
    archivePrefix = "arXiv",
    primaryClass = "hep-ph",
    doi = "10.1140/epjc/s10052-021-08968-2",
    journal = "Eur. Phys. J. C",
    volume = "81",
    number = "2",
    pages = "181",
    year = "2021"
}

@article{Murphy:2020rsh,
    author = "Murphy, Christopher W.",
    title = "{Dimension-8 operators in the Standard Model Effective Field Theory}",
    eprint = "2005.00059",
    archivePrefix = "arXiv",
    primaryClass = "hep-ph",
    doi = "10.1007/JHEP10(2020)174",
    journal = "JHEP",
    volume = "10",
    pages = "174",
    year = "2020"
}

@article{Murphy:2020cly,
    author = "Murphy, Christopher W.",
    title = "{Low-Energy Effective Field Theory below the Electroweak Scale: Dimension-8 Operators}",
    eprint = "2012.13291",
    archivePrefix = "arXiv",
    primaryClass = "hep-ph",
    doi = "10.1007/JHEP04(2021)101",
    journal = "JHEP",
    volume = "04",
    pages = "101",
    year = "2021"
}

@article{Li:2020tsi,
    author = "Li, Hao-Lin and Ren, Zhe and Xiao, Ming-Lei and Yu, Jiang-Hao and Zheng, Yu-Hui",
    title = "{Low energy effective field theory operator basis at d \ensuremath{\leq} 9}",
    eprint = "2012.09188",
    archivePrefix = "arXiv",
    primaryClass = "hep-ph",
    doi = "10.1007/JHEP06(2021)138",
    journal = "JHEP",
    volume = "06",
    pages = "138",
    year = "2021"
}

@article{Li_2021,
   title={Complete set of dimension-eight operators in the standard model effective field theory},
   volume={104},
   ISSN={2470-0029},
   url={http://dx.doi.org/10.1103/PhysRevD.104.015026},
   DOI={10.1103/physrevd.104.015026},
   number={1},
   journal={Physical Review D},
   publisher={American Physical Society (APS)},
   author={Li, Hao-Lin and Ren, Zhe and Shu, Jing and Xiao, Ming-Lei and Yu, Jiang-Hao and Zheng, Yu-Hui},
   year={2021},
   month=jul }

@article{Li:2021tsq,
    author = "Li, Hao-Lin and Ren, Zhe and Xiao, Ming-Lei and Yu, Jiang-Hao and Zheng, Yu-Hui",
    title = "{Operator bases in effective field theories with sterile neutrinos: d \ensuremath{\leq} 9}",
    eprint = "2105.09329",
    archivePrefix = "arXiv",
    primaryClass = "hep-ph",
    doi = "10.1007/JHEP11(2021)003",
    journal = "JHEP",
    volume = "11",
    pages = "003",
    year = "2021"
}

@article{Li:2022tec,
    author = "Li, Hao-Lin and Ren, Zhe and Xiao, Ming-Lei and Yu, Jiang-Hao and Zheng, Yu-Hui",
    title = "{Operators for generic effective field theory at any dimension: on-shell amplitude basis construction}",
    eprint = "2201.04639",
    archivePrefix = "arXiv",
    primaryClass = "hep-ph",
    doi = "10.1007/JHEP04(2022)140",
    journal = "JHEP",
    volume = "04",
    pages = "140",
    year = "2022"
}

@article{Harlander:2023ozs,
    author = "Harlander, Robert V. and Schaaf, Magnus C.",
    title = "{AutoEFT: Automated operator construction for effective field theories}",
    eprint = "2309.15783",
    archivePrefix = "arXiv",
    primaryClass = "hep-ph",
    reportNumber = "TTK-23-25; P3H-23-066",
    doi = "10.1016/j.cpc.2024.109198",
    journal = "Comput. Phys. Commun.",
    volume = "300",
    pages = "109198",
    year = "2024"
}

@article{Georgi:1991ch,
    author = "Georgi, Howard",
    title = "{On-shell effective field theory}",
    reportNumber = "HUTP-91-A014",
    doi = "10.1016/0550-3213(91)90244-R",
    journal = "Nucl. Phys. B",
    volume = "361",
    pages = "339--350",
    year = "1991"
}

@article{Li:2023edf,
    author = "Li, Xu and Zhou, Shun",
    title = "{One-loop Matching and Running via On-shell Amplitudes}",
    eprint = "2309.10851",
    archivePrefix = "arXiv",
    primaryClass = "hep-ph",
    month = "9",
    year = "2023"
}

@article{DeAngelis:2023bmd,
    author = "De Angelis, Stefano and Durieux, Gauthier",
    title = "{EFT matching from analyticity and unitarity}",
    eprint = "2308.00035",
    archivePrefix = "arXiv",
    primaryClass = "hep-ph",
    reportNumber = "CERN-TH-2023-150",
    doi = "10.21468/SciPostPhys.16.3.071",
    journal = "SciPost Phys.",
    volume = "16",
    pages = "071",
    year = "2024"
}

\end{document}